\documentclass[review,times]{elsarticle}
\usepackage{lineno,hyperref}
\modulolinenumbers[5]
\usepackage[labelfont=bf,justification=raggedright,singlelinecheck=false]{caption}
\usepackage{graphicx}
\usepackage{booktabs}
\usepackage[english]{babel}
\usepackage[utf8]{inputenc}
\RequirePackage[T1]{fontenc}
\usepackage{babel}
\usepackage{color,soulutf8}
\usepackage{amsmath, amssymb}
\usepackage{multirow}
\usepackage{ragged2e}
\usepackage[flushleft]{threeparttablex}
\usepackage{rotating}
\usepackage{adjustbox}
\usepackage{amsmath}
\usepackage{latexsym}
\usepackage{hyperref}
\usepackage{fullpage}
\usepackage{longtable}
\usepackage{comment}
\usepackage{pdflscape}
\usepackage{float}
\usepackage{rotfloat}
\usepackage{setspace}
\journal{}

\biboptions{authoryear}

\begin{document}

\begin{frontmatter}

\title{Portfolio optimization based on forecasting models using vine copulas: An empirical assessment for the financial crisis\tnoteref{t1}}
\tnotetext[t1]{We are grateful to participants at the Financial Econometrics Workshop at Örebro University (Sweden), 7-8 November  2018, the 12th International Conference on
Computational and Financial Econometrics (CFE 2018) in Pisa/Italy,  14-16 December 2018, the Emerging Topics in Financial Economics Workshop at Linköping University (Sweden), 21 February 2019, the  Infiniti Conference in Glasgow (Scotland), 11-12 June 2019, the Econometrics and Statistics Conference 24-26 June 2019 in Taichung (Taiwan), and the Vine Copula Workshop at TU Munich, 8-9 July 2019, for helpful comments and suggestions. The usual disclaimer applies.}

\author[mymainaddress]{Maziar Sahamkhadam}

\author[mysecondaryaddress]{Andreas Stephan\corref{mycorrespondingauthor}}
\cortext[mycorrespondingauthor]{Corresponding author}
\ead{andreas.stephan@ju.se}

\address[mymainaddress]{Linnaeus University, V\"{a}xj\"{o}, Sweden}
\address[mysecondaryaddress]{J\"{o}nk\"{o}ping International Business School, J\"{o}nk\"{o}ping University, Sweden}

\begin{abstract}
We employ and examine vine copulas in modeling symmetric and asymmetric dependency structures and forecasting financial returns. We analyze the asset allocations performed during the 2008--2009 financial crisis and test different portfolio strategies such as maximum Sharpe ratio, minimum variance, and minimum conditional Value-at-Risk. We then specify the regular, drawable, and canonical vine copulas, such as the Student$-t$, Clayton, Frank, Joe, Gumbel, and mixed copulas, and analyze both in-sample and out-of-sample portfolio performances. Out-of-sample portfolio back-testing shows that vine copulas reduce portfolio risk better than simple copulas.
Our econometric analysis of the outcomes of the various models shows that in terms of reducing conditional Value-at-Risk, D-vines appear to be better than R- and C-vines. Overall, we find that the Student$-t$ drawable vine copula models perform best with regard to risk reduction, both for the entire period 2005--2012 as well as during the financial crisis.
\end{abstract}


\end{frontmatter}

\setcounter{page}{0}
\newpage

\section{Introduction}

During a financial crisis, asset returns show different behaviors compared to those seen during non-crisis periods \citep{ZHANG2014}. This includes the tail dependence between assets. Tail dependence in financial returns leads to different distributional assumptions that can be used in an investor's utility function maximization \citep{FERNANDEZ2005}. Particularly, in downside risk minimization, lower tail dependence can affect the performance of portfolio strategies. In addressing this issue, Sklar's copula theory has gained popularity in modeling the dependence structure of financial returns, both symmetric and asymmetric \citep{BENSAIDA2018,frey2003dependent, li2000default, patton2006modelling}. The variety of copula families enables researchers and investors to estimate the joint distribution of returns with properties ranging from lower to upper tail dependence \citep{joe1997multivariate}. Moreover, pair-copula construction (PCC) and vine copulas allow for use of different copula families to estimate the asset returns' dependence structure, leading to a more flexible modeling \citep{aas2009pair}.

In portfolio allocation techniques, minimum risk and maximum reward-to-risk methods can capture the investor's general purpose of adopting a certain portfolio strategy. When using the latter method, both classical volatility and downside risk can measure the portfolio risk. These optimization methods can be combined with forecasting models that, in general, create expectations on returns' properties, e.g., mean, volatility and tail dependence \citep{RIGHI2013}. Therefore, appropriate distributional assumptions for utility function maximization, including symmetric or asymmetric, can lead to uncertainty in utilizing the forecasting models during a financial crisis. The above arguments raise some general questions. Can the copula families that are sensitive to lower tail offer a more reliable forecasting model and thus a well-performed downside risk-based portfolio strategy during a crisis period? Do the asymmetric assumptions for return distribution improve the maximum reward-to-risk ratio portfolio optimization? Do the financial returns' asymmetry and dependence structure change over the evolution of financial crisis? 

This study contributes to the existing literature by addressing these issues. First, we examine both asymmetric and symmetric copula families in a portfolio optimization and back-testing setting. Focussing on asset allocation, we compare the performance of these copula families during the period of the 2008--2009 global financial crisis. Considering different copula families, we apply several forecasting models to both in-sample and out-of-sample stock market estimation. R-vine, D-vine, and C-vine copula structures are specified; these include the Student$-t$ (symmetric upper and lower tail dependence), Clayton (captures asymmetry and lower tail), Frank (captures symmetry), Joe (sensitive to upper tail), Gumbel (sensitive to upper tail), and mixed (a mixture of all families) copulas. We optimize the portfolios, including maximum reward-to-risk ratio (the Sharpe ratio), global minimum variance (GMV), and minimum conditional Value-at-Risk (CVaR), for simulated returns from each copula distribution. Using portfolio-related measures, we compare the performance of the estimated copula families, both symmetric and asymmetric, in maximizing the investor's utility function. As for the copula families' performance in forecasting portfolio downside risk, we use the common value-at-risk (VaR) and expected shortfall (ES) back-testing procedures. Second, not only the long-term, but also short-term investments are considered. We create and compare 2-year holding periods over the targeted sample and estimate portfolio measures that capture the utility function maximization, including conditional Value-at-Risk (CVaR), Sharpe ratio (SR) and standard deviation. This allows for a better understanding of the copula-based portfolio strategies' performance over a short-term investment horizon during the global financial crisis. Finally, using these measures, we perform a regression analysis on the effect of copula families and vine structures on the portfolio out-of-sample performance.


Out-of-sample portfolio back-testing shows that vine copulas are better at reducing portfolio risk than simple copulas. In the case of portfolio out-of-sample CVaR, Frank and Student$-t$ vine copulas result in lower downside risk. Our short-term investment analyses reveal gains obtained from vine copula-based SR portfolios. Although most of the model fails to pass the downside risk back-testing, asymmetric copulas (e.g., Clayton) are better able to forecast the downside risk. Copula families capturing no tail dependence (Frank) and upper tail dependence (Joe and Gumbel) lead to higher terminal portfolio values over the financial crisis. The econometric analysis of the target measures from the various optimal portfolio models reveals that D-vines, in general, reduce downside risk measured by CVaR better than R- or C-vines, and that the Student-$t$ D-vine copula is overall best both for the entire period and during the financial crisis. However, the simple Gumbel copula model shows better effects regarding increasing the Sharpe ratio in both periods. Somewhat surprisingly, mixed vines do not show better performance in terms of risk reduction compared to the less flexible single family vine copula models.

The remainder of this paper is structured as follows. Section \ref{LR} presents a short review of the previous studies. Empirical methods such as copula modeling and portfolio optimization are discussed in Section \ref{Emp.Met.}. The data used are presented in Section \ref{Data}. Section \ref{Emp.An.} presents our empirical analysis. Finally, Section \ref{Conclusion} discusses the main conclusions of the paper.

\section{Literature Review}\label{LR}
In this study, we use copula modeling jointly with portfolio optimization and examine the various copula families with different structures for modeling symmetric and asymmetric tail dependence. To this end, we first review the literature, focussing on studies using vine copulas in either portfolio allocation or downside risk modeling.

\cite{deng2011portfolio} utilize C-vine and D-vine copula structures for mean-CVaR allocation. The study compares the vine copula structures with the Student$-t$ copula in an in-sample simulation study, to show that vine copulas provide a more efficient frontier. \cite{de2012choosing} propose a robust pair-copula, the Markowitz mean-variance optimization, with several rebalancing strategies. They conclude that the suggested allocation method increases the portfolio returns in long-term investment with lower turnover. \cite{weiss2013forecasting} utilize D-vine copula to model the dependence structure of intraday bid-ask spreads and stock portfolio, and conclude that the D-vine copula is appropriate for strong tail dependence. \cite{low2013canonical} examine the canonical vine copula structure and compare the Min-CVaR portfolio strategies obtained using several copula distributions, such as Gaussian, Student$-t$, and Clayton. Clayton is the only copula family they specified in the vine structure. They find better results by applying asymmetric tail dependence modeling to the Min-CVaR portfolio strategy. \cite{hernandez2014oil} evaluates the C-vine and D-vine copula structures for coal-uranium and oil-gas stocks, to conclude that the C-vine copula structure is more suitable for modeling the tail dependence of energy stocks. In addition, the author shows that C-vine-based asset allocation out-performs the classical nonlinear optimization. \cite{zhang2014forecasting} test different vine copula structures in VaR and CVaR back-testing for the daily international stock market returns, to find the D-vine copula structure superior to the C-vine and R-vine structures. \cite{reboredo2015vine} estimate the D-vine dependence structure for modeling the downside risk (VaR and CoVaR) of MSCI and bond indexes. They use the different copula families that capture no tail dependence, symmetric and asymmetric tail dependence, and CoVaR as a measure of systemic risk, to find that the systemic effect of sovereign debt is different before, during, and after a crisis. \cite{reboredo2015downside} investigate the spillover between four commodity (precious metal) returns. They test the VaR and CoVaR modeling obtained with D-vine copula for weekly data, to find asymmetric downside and upside spillover effects. \cite{siburg2015forecasting} suggest nonparametric tail dependence estimation based on C-vine copula, examined and selected over D-vine, and show that their model out-performs parametric estimation in out-of-sample VaR back-testing. \cite{righi2015forecasting} apply serial PCC and D-vine copulas to VaR and CVaR estimation in four stock markets, and find significant variance in dependence structure owing to different lags. \cite{weiss2015mixture} suggest a mixed PCC for VaR forecasting, and compare the performance of their suggested mixture copula with common vine copulas.
\cite{bekiros2015multivariate} perform vine copula-based minimum risk allocation for mining stock portfolios during a financial crisis and show the performance of vine copula in forecasting tail dependence. \cite{aloui2016relationship} study the tail dependence between crude oil, stock markets, and the currency trade-weighted index by applying vine copulas, to find better VaR forecasts. \cite{sukcharoen2017hedging} investigate the downside risk hedging strategies of oil refineries based on vine copulas and conclude that vine copula structures, particularly the D-vine structure, outperform nonparametric and standard multivariate copula models in out-of-sample hedging performance. \cite{uddin2018multivariate} use C-vine in modeling tail dependence and spillover effects for energy commodities and propose different portfolio strategies. \cite{yu2018measuring} propose a mixed R-vine copula for modeling the downside risk (VaR and CVaR) of crude oil and compare this mixed version with the Joe and Gumbel copula families. VaR and CVaR back-testing shows better results from mixed R-vine models.

From the above literature review, most of the study results are data related, where the vine copula structure (e.g., D-vine, C-vine, or R-vine) selection is based on the targeted dataset. A comparison of the results obtained with the different vine copula structures can enhance the reliability of the forecasting models. A study focusing on the portfolio optimization techniques used during a financial crisis can expand the application and usage of vine copulas in financial analysis. Thus, an in-depth investigation of the different copula modeling with the different distributional properties of stock markets during a financial crisis would be an invaluable contribution to the existing literature.

\section{Empirical Methods}\label{Emp.Met.}
In this section, we review copula theory, the C-vine, D-vine, and R-vine structures, and their estimation methods. We then continue with our marginal modeling and portfolio optimization methods.

\subsection{Pair Copula Construction}\label{PCC}
As suggested by \cite{sklar1959fonctions}, copulas can be used to estimate the joint distribution of univariate marginals. A combination of the marginal distribution and a copula function based on the dependence structure of assets results in a multivariate (joint) distribution. This joint distribution can be utilized to model the financial returns based on separately estimated marginals and the dependence structure. According to \cite{sklar1959fonctions}, a $d$-dimensional distribution function $F$ can be estimated from marginal distributions $F_1,...,F_d$ and a $d$-dimensional copula $C$. For a bivariate copula, we have

\begin{equation}\label{BiCop}
\left \{
\begin{array}{ll}
\forall \textbf{z} \in \Re^{d}:F(z_1,z_2)=C(F_1(z_1),F_2(z_2))=C(u_1,u_2)\\
\forall \textbf{u} \in [0,1]^d:z_n=F_n^{-1}(u_n)\\
C(u_1,u_2)=F(F_1^{-1}(u_1),F_2^{-1}(u_2))=F(z_1,z_2)
\end{array}
\right.
\end{equation}

Assuming reversibility of the copula function $C(u_1,u_2)$, \cite{sklar1973random} suggests differentiability of the marginal distributions $F_n$ and copula functions $C$. Therefore, the joint density of multivariate distribution is a product of the marginal densities $f_n(z_n)$ and copula density $c(u_1,u_2)$. In this case, the joint and copula densities are, respectively,

\begin{equation}\label{BiCopJoint}
\left \{
\begin{array}{ll}
f(z_1,z_2)=f_1(z_1)\times f_2(z_2)\times c[F_1(z_1),F_2(z_2)]\\
\\
c(u_1,u_2)=\frac{\partial^dC(u_1,u_2)}{\partial u_1 \partial u_2}
\end{array}
\right.
\end{equation}

The bivariate representation of the copula function and its relation with marginal distribution can be extended to a different formulation (families) of the copula used to estimate the dependence structure. Two main copula categories are the elliptical and Archimedean copula families. Examples of the elliptical families are Gaussian and Student$-t$. The latter one captures symmetric tail dependence. By using a generator function ($\varphi$), Archimedean copulas can provide more flexibility in modeling asymmetric dependence for both upper and lower tail. Examples of Archimedean families are Clayton (asymmetric lower tail); Joe; Gumbel; BB6 and BB8 (asymmetric upper tail); BB1 and BB4 (asymmetric lower and upper tails); and Frank (symmetric no tail dependence). In addition, rotated versions of the Archimedean families can be used owing to the parameter restrictions imposed by generator functions of the non-rotated families (see \cite{brechmann2013modeling} for more details on rotated copula families). Table \ref*{Cop.Fam} presents the bivariate copula function $C(u_1,u_2)$, generator $\varphi(u)$, lower tail $\lambda_L$, and upper tail $\lambda_U$ dependence for different families.

\begin{landscape}
\begin{figure}[h]
	\centering
	\caption{Examples of $5$-dimensional Vine Trees}\label{Fig1}
	\includegraphics[width=17cm]{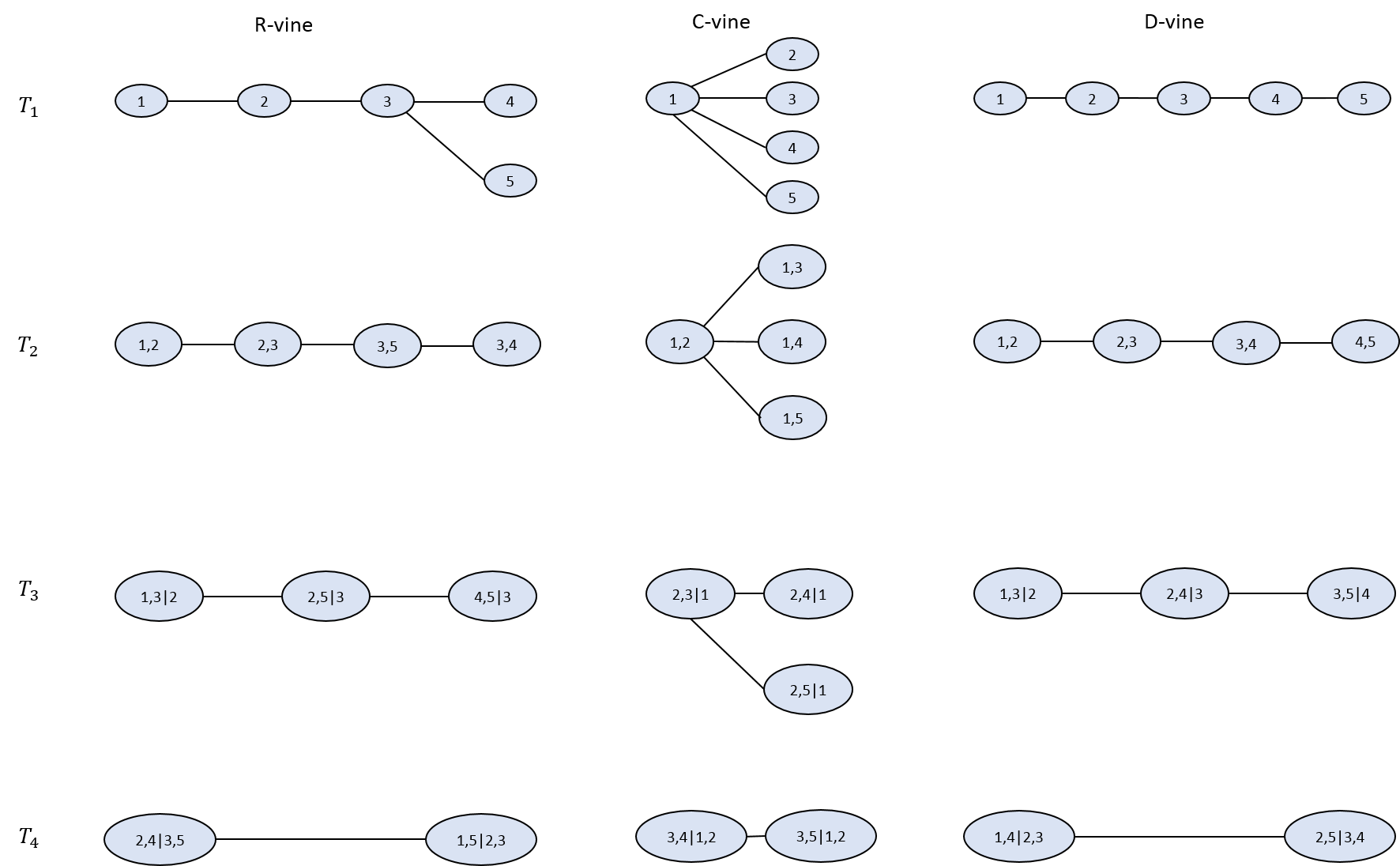}
\end{figure}
\end{landscape}

Some of the bivariate copulas (e.g., ~Student$-t$) can be extended to estimate the multivariate dependence structure, but there are limitations to capturing a multi-parameter dependence structure. To overcome these limitations, \cite{aas2009pair} suggest PCCs based on \cite{joe1997multivariate}, which estimates the marginal conditional distribution function, \cite{bedford2002vines}, which considers the regular vine (R-vine) in multivariate statistical modeling, and \cite{kurowicka2004distribution}, which introduces canonical (C-vine) and drawable (D-vine) structures. \cite{aas2009pair} combine the marginal conditional distribution with vines, which are graphical representations of the dependence structure. The vine approach for a $d$-dimensional PCC involves $d(d-1)/2$ pair-copulas and $d-1$ linked trees. When using bivariate copulas, the first vine tree consists of the dependence of one variable (the first root node) for each pair and the conditional dependence of other variables (the second root node, third root node, and so on). In other words, for each tree model, once the first root node is modeled, the second root node is based on the first one, and the third root node is based on the second one. This approach is called recursive conditioning. Figure \ref{Fig1} illustrates the tree structure for $5$-dimensional R-vine, C-vine, and D-vine copulas.

For the regular vine copula, the joint density function is defined as \citep{dissmann2013selecting}:
\begin{equation}\label{RVine}
f^{Rvine}(\textbf{z})=\prod_{j=1}^d f_j(z_j)\times\prod_{i=1}^{d-1}\prod_{e\in E_i}c_{C_{e,\alpha},C_{e,b}|D_{e}}(F_{C_{e,\alpha}|D_e}(z_{C_{e,\alpha}}|\textbf{z}_{D_e}),\\F_{C_{e,b}|D_e}(z_{C_{e,b}}|\textbf{z}_{D_e}))
\end{equation}
where $e=\{\alpha,b\}$, and $\textbf{z}_{D_e}$ denotes the variables in $D_e$, that is, $\textbf{z}_{D_e}=\{x_i|i\in D_e\}$. $f_j$ is the density of $F_j$ for $j=1,...,d$ and $\textbf{z}=(z1,...,z_d)$.

The joint density function based on a $d$-dimensional canonical vine copula (C-vine) is \citep{brechmann2013modeling}:
\begin{equation}\label{CVine}
	f^{Cvine}(\textbf{z})=\prod_{j=1}^d f_j(z_j)\times\prod_{i=1}^{d-1}\prod_{n=1}^{d-i}c_{i,i+n|1:(i-1)}(F(z_i|z_1,...,z_{i-1}),\\F(z_{i+n}|z_1,...,z_{i-1})|\boldsymbol\Omega_{i,i+n|1:(i-1)})
\end{equation}
where $f_j$ and $c_{i,i+n|1:(i-1)}$ are respectively the marginal densities and bivariate copula densities, and $\boldsymbol\Omega_{i,i+n|1:(i-1)}$ and $i=1,...,d-1$ denote respectively the parameters and root nodes. However, for the drawable vine copula (D-vine), the joint density function is \citep{brechmann2013modeling}:
\begin{equation}\label{DVine}
	f^{Dvine}(\textbf{z})=\prod_{j=1}^d f_j(z_j)\times\prod_{i=1}^{d-1}\prod_{n=1}^{d-i}c_{n,n+i|(n+1):(n+i-1)}(F(z_n|z_{n+1},...,z_{n+i-1}),\\F(z_{n+i}|z_{n+1},...,z_{n+i-1})|\boldsymbol\Omega_{n,n+i|(n+1):(n+i-1)}).
\end{equation}

\subsection{Vine Copula Estimation}\label{MLE}
\cite{joe1997multivariate} show how to solve a $d-$dimensional conditional distribution function:
\begin{equation}\label{JoeVine}
	F_{C_{e,\alpha}|D_e}(z_{C_{e,\alpha}}|\textbf{z}_{D_e})=\frac{\partial C_{C_\alpha|D_\alpha}(F_{C_{\alpha,\alpha_1}|D_\alpha}(z_{C_{\alpha,\alpha_1}}|\textbf{z}_{D_\alpha}),F_{C_{\alpha,\alpha_2}|D_\alpha}(z_{C_{\alpha,\alpha_2}}|\textbf{z}_{D_\alpha}))}{\partial F_{C_{\alpha,\alpha_2}|D_\alpha}(z_{C_{\alpha,\alpha_2}}|\textbf{z}_{D_\alpha})},
\end{equation}
where, $E_i\ni e=\{\alpha,b\},\alpha=\{\alpha_1,\alpha_2\}$.

The log-likelihood function for C-vine  and D-vine copulas are \citep{brechmann2013modeling}:
\begin{equation}\label{CVinemle}
	\ell^{Cvine}(\boldsymbol\Omega|\textbf{z})=\sum_{j=1}^d\sum_{i=1}^{d-1}\sum_{n=1}^{d-i}log[C_{i,i+n|1:(i-1)}(F_{i|1:(i-1)},\\F_{i+n|1:(i-1)}|\boldsymbol\Omega_{i,i+n|1:(i-1)})]
\end{equation}

\begin{equation}\label{DVinemle}
	\ell^{Dvine}(\boldsymbol\Omega|\textbf{z})=\sum_{j=1}^d\sum_{i=1}^{d-1}\sum_{n=1}^{d-i}log[C_{n,j+i|(n+1):(n+i-1)}(F_{n|(n+1):(n+i-1)},\\F_{n+i|(n+1):(n+i-1)}|\boldsymbol\Omega_{n,n+i|(n+1):(n+i-1)})].
\end{equation}
The log-likelihood function for the R-vine copula is
\begin{equation}\label{RVinemle}
	\ell^{Rvine}(\textbf{z}_{D_e}|\textbf{z})=\sum_{j=1}^d\sum_{i=1}^{d-1}\sum_{e\in E_i}log[C_{C_{e,\alpha},C_{e,b}|D_{e}}(F_{C_{e,\alpha}|D_e}(z_{C_{e,\alpha}}|\textbf{z}_{D_e}),\\F_{C_{e.b}|D_e}(z_{C_{e,b}}|\textbf{z}_{D_e})].
\end{equation}

\subsection{Marginal Modeling}\label{MarginalModeling}
As shown above, we use copulas to estimate the joint distribution of univariate marginals. To model the univariate marginals that can be utilized as inputs to copula, we use the autoregressive process in the mean equation and standard generalized autoregressive conditional heteroskedasticity model in the volatility equation. We do not evaluate the different forecasting models in terms of mean and volatility equations, but focus on the competing models based on different vine copula structures and families. Therefore, the forecasting models have two main steps. First, we fit the AR-GARCH model and obtain one-step-ahead mean and volatility forecasts. We then filter out the standardized residuals and use probability integral transformation to obtain marginal uniforms. In the second step, we fit a vine copula to these uniforms, estimate the dependence structure, and simulate one-step ahead returns from the estimated joint distribution. The marginal modeling used in both steps can be defined as

\begin{equation}\label{AR-GARCH}
\left \{
\begin{array}{ll}
r_{jt}=\mu_{j}+\phi_{j}r_{n,t-1}+\epsilon_{jt}\\
\epsilon_{jt}=z_{jt}\;\sqrt{h_{jt}}\\
z_{jt}\sim \text{skewed Student}-t (\eta_j,\zeta_j)\\
h_{jt}=\omega_j+\alpha_{1j}\epsilon_{j,t-1}^2+\beta_{1j}h_{j,t-1}
\end{array}
\right.
\end{equation}
where $r_{jt}$ is the returns for asset $j=1,2,...,d$, and $z_{jt}$ is the standardized residuals, with parameter restrictions $\omega_j>0$, $\alpha_{1j}\geq0$, $\beta_{1j}\geq0$, $\alpha_{1j}+\beta_{1j}<1$, $\phi_j\neq 0$, $2<\eta_j<\infty$, and $-1<\zeta_j<1$. The conditional distribution in the AR-GARCH model is skewed Student$-t$.


\subsection{Utility Function and Portfolio Allocation}\label{AllocationMethod}
Since we are examining the vine copula structures and families in a portfolio optimization setting, we need to allocate for different investor utility functions. This study focuses on the 2008--2009 financial crisis period when the investor must have sought for lower portfolio risk. As suggested by \cite{markowitz1952portfolio}, volatility can be considered a measure of portfolio risk. The portfolio volatility can be minimized with or without constraining the portfolio target return, to obtain the Markowitz mean-variance and global MV strategies. The minimization in both portfolio strategies involves quadratic programing. However, portfolio downside risk is an appropriate measure to capture the portfolio tail risk and losses. During a financial crisis, minimizing the portfolio downside risk would reduce the investor's losses and lead to a more secure investment strategy for risk-averse investors. CVaR is widely used as a measure of portfolio downside risk \citep{low2013canonical, reboredo2015downside, righi2015forecasting}. Similar to portfolio volatility, to minimize CVaR, a portfolio target return can be added as a constraint leading to a mean-CVaR strategy. However, we focus on the Min-CVaR strategy without the target return constraint as optimization would only reduce the downside risk. As regards the Min-CVaR strategy, the optimization problem can be solved with linear programing \cite{rockafellar2000optimization}. In addition to the MV and Min-CVaR strategies, which capture and reduce the investor's risk, we can test for Sharpe ratio (SR) maximization, where the portfolio return is maximized over the volatility \citep{sharpe1994sharpe}. This strategy would be utilized by risk-neutral investors who look for higher portfolio returns rather than both the MV and Min-CVaR strategies. Although increasing returns would lead to higher volatility, the investor will be willing to bear higher risk for a higher return. Particularly, during a financial crisis, the investor would prefer to gain from her investment strategy if plausible. Hence, we focus on the MV, CVaR, and SR portfolio optimization strategies and evaluate the effects of the symmetric and asymmetric dependence structure estimations with vine copulas on each allocation strategy (see \cite{sahamkhadam2018portfolio} for more details on MV, CVaR, and SR portfolio strategies).
\newpage
\section{Financial Data}\label{Data}
We examine the performance of different portfolio strategies based on vine copula models using the logarithmic returns of 12 international markets: the United States (S\&P 500), the United Kingdom (FTSE 100), Germany (DAX 30), Spain (IBEX 35), South Korea (KOSPI), Japan (TOPIX), Canada (S\&P/TSX), Sweden (OMXS 30), Switzerland (SMI), Finland (OMXH), Argentina (MERVAL), and Egypt (HRMS). The sample period is from June 3, 2003, to December 12, 2012, with 2500 daily returns for each stock market. This period is chosen because it fully captures the performance of an investment strategy during the 2008--2009 global financial crisis as well as the influence of a training sample from before to after the crisis. In this case, we also investigate how fast our portfolio strategies can recover from the financial crisis.

Table \ref{Desc.Stat} presents the descriptive statistics of stock market returns for two sample periods including the global financial crisis (Panel A: 2008-2009) and the full sample (Panel B: 2003-2012). During the crisis, all markets show negative average return, except for Argentina. The lowest average return is reported for Finland. In the table, international stock markets show different volatilities during the 2008--2009 financial crisis, ranging from 1.77\% for Switzerland to 2.51\% for Argentina. The lowest (highest) minimum return is reported for Egypt (Germany), while Japan has the maximum return over the crisis. Germany, Spain, Sweden, Switzerland and Finland show positive skewness. The positive kurtosis reported for all the series is an indicator of empirical distribution peaking more than the Gaussian distribution. The significant Jarque-Bera normality test results suggest non-Gaussian empirical distribution for all markets over the global financial crisis. Finally, the 10\% VaR and CVaR for each market shows the variability of losses due to the financial crisis. In this sense, the top three markets suitable for investment during the 2008--2009 crisis were Switzerland, United Kingdom and Spain, being less affected in terms of downside risk. These measures are important as it enables the investor to compare the performance of a diversified portfolio strategy with investment in one of these markets.

\section{Empirical Analysis}\label{Emp.An.}
We begin our empirical investigation with an in-sample estimation of the forecasting models. This gives a better understanding of parameter estimation and the general expectations of the forecasting models. First, we analyze the AR-GARCH parameter estimation. We then examine vine copula estimation. As the final in-sample analysis, we consider the risk--return relationship of the SR and CVaR portfolio strategies as an efficient frontier. We will continue our empirical analysis by comparing the out-of-sample performance of several portfolio strategies obtained using different vine copula structures. Finally, we perform VaR back-testing.

\subsection{AR-GARCH Estimation}\label{AR.GARCH.ES}
To estimate the dependency structure of stock markets, we first need to model the univariate marginals. As mentioned in Section \ref{MarginalModeling}, the marginals are assumed to follow an AR-GARCH process, where a constant ($\mu_j$) and the autoregressive term ($\phi_{j}r_{j,t-1}$) with one lag constitute the conditional mean. The volatility of this process consists of a standard GARCH(1,1), with standardized residuals ($z_j$) modeled using skewed Student$-t$ distribution. Table \ref{AR.GARCH.Param} reports the parameter estimation. As the table shows, during the financial crisis (Panel A), the constant terms in both mean and volatility equations ($\mu_j$ and $\omega_j$) are not significant in almost all cases. All the GARCH parameters ($\alpha_{1j}$ and $\beta_{1j}$) are significant. The shape and skewness parameters reported in the table indicate non-Gaussian distribution for all assets, except for Sweden, during the global financial crisis. On the other hand, over the full sample (Panel B), all the parameters are significant at 1\% level, except for AR term ($\phi_j$) in some cases.

\subsection{Vine Copula}\label{VineCopula}
Having obtained the standardized residuals ($z_j$), we estimate the dependence structure of the stock markets. To compare the different vine structures, we use the mixed versions of the C-vine, D-vine, and R-vine copulas \citep{Czado2019}. In these mixed versions, we allow for selection of pairwise copulas from all the families listed in Table \ref{Cop.Fam} based on the sequential estimations suggested in \cite{brechmann2013modeling} and \cite{dissmann2013selecting}. We then use the selected families, with the estimated parameters, as the initial values for vine copula maximum likelihood estimation.\footnote{Note that before the sequential estimation, we need to characterize the dependency structure by choosing the order of variables in the vine structures. We follow \cite{low2013canonical} and define the order of assets according to their total correlation with other assets. We note that this order would change if we use the rolling window procedure.}

\begin{landscape}
\begin{figure}[H]
	\centering
	\caption{Vine Copula Structures for International Markets}\label{Fig2}
	\includegraphics[height=0.5\textheight]{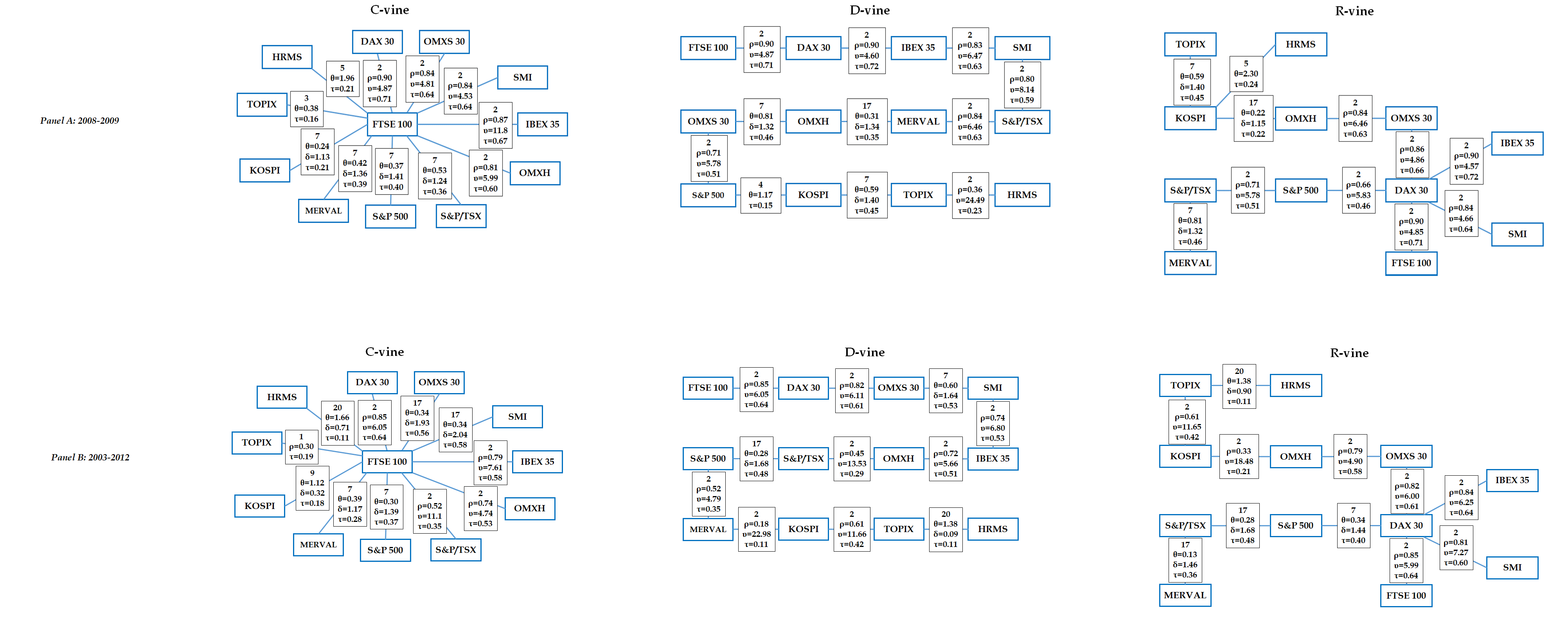}
	\caption*{Notes: This figure presents the vine copula (C-vine, D-vine, and R-vine) estimation results for stock market marginal uniforms $u_n$ obtained using the AR-GARCH process with skewed Student$-t$ conditional distribution. In Panel A, the sample period is from January 1, 2008 to December 31, 2009. Panel B reports the result from June 3, 2003 to December 31, 2012. The first tree is shown for each vine copula. The copula families (as numbers), parameters ($\rho$ and $\nu$ for elliptical families and $\theta$ and $\delta$ for Archimedean families), and Kendal's $\tau$ are reported at the edges. See Table \ref{BiCop} for more information on copula families.}
\end{figure}
\end{landscape}

Figure \ref*{Fig2} shows the different vine structures used for the whole sample, along with the selected bivariate copula families. As mentioned in Section \ref*{PCC}, a $12-$dimensional dataset has 11 tree structures and 66 pair copulas as nodes. Here, we show only the first trees for the C-vine, D-vine, and R-vine copulas. From the figure, FTSE 100 has the highest correlation with other markets, and is at the first node. From the figure, copula selection and estimation is different for the two periods. For instance, in tree 1 for the C-vine copula during the financial crisis (Panel A), the dependency between United Kingdom and Sweden is modeled based on Student$-t$ with Kendal's $\tau$  estimated as 0.64. However, during the whole sample (Panel B), rotated BB1 is selected with $\tau=0.56$.

The vine structure plots give an understanding of how mixed copula versions are constructed and what copula families are included, along with their estimated parameters. However, they do not provide the appropriateness of copula families nor the vine structure. Therefore, a goodness-of-fit procedure is required to approximate the best copula model. In doing so, we use empirical copula process (ECP) tests and estimate Cramer-von Mises (CvM) and Kolmogorov-Smirnov (KS) $t-$statistics \citep[see][for more details on the tests]{genest2009goodness,schepsmeier2015efficient}. According to Table \ref{GOF}, Student$-t$ and Frank copulas have the best fit during the global financial crisis. Regrading the whole sample, the best copulas are Student$-t$ and Joe. Considering these results, symmetric tail dependency exists for both periods and is captured by Student$-t$ copula. In terms of vine structure, in both sample periods, canonical vine has the best fit compared to drawable and regular vines.

\subsection{Portfolio In-Sample Performance}\label{InS.Per}
To examine the effect of copula models on portfolio performance, we use in-sample estimation of asset returns. In doing so, we use the targeted sample to fit the AR-GARCH model and several copula models. We simulate 10000 returns from the corresponding AR-GARCH-Copula process and perform portfolio optimization. Finally, we use the asset returns over the targeted sample and the estimated weights to construct portfolio strategies and obtain performance measures. We use equally-weighted (EQW) and historical portfolios as the benchmarks.

Since, all the models use the same AR-GARCH process to estimate conditional mean and volatilities, the in-sample performance measures can help to understand the impact of copula models on portfolio strategies. Tables \ref{InS.SR}-\ref{InS.GMV} present the results. Regarding SR portfolios (maximum Sharpe ratio), all of the forecasting models give negative returns during the global financial crisis (Panel A). However, there are some gain from vine copulas in terms of average return during this period. For instance, Dvine copulas give the lowest negative average return, in particular, Gumbel family (with a mean return of -0.044). At the same time, these models result in higher volatility and downside risk. Clayton and Student$-t$  Rvines show better results in terms of the portfolio downside risk (CVaR), which is due to their sensitivity to lower tail. In almost all cases, Cvine and Dvine structures show small improvement for portfolio economic performance (terminal wealth). For CVaR portfolios (minimum CVaR), Student$-t$ copulas show slight improvement in portfolio average return during the crisis. Similar to SR portfolios, Clayton and Student$-t$ Rvine copulas provide lower portfolio CVaR, comparing to other copula models. Furthermore, mixed, Student$-t$ and Gumbel vines have the lowest portfolio volatility for GMV strategy, in both sample periods.

In general, the results of in-sample performance indicate more gain obtained from copula-based SR portfolios. However, in most cases, there is no gain from copulas for CVaR and GMV. In particular, during the global financial crisis, the utility function maximization is better obtained from historical allocation.

\subsection{Portfolio Out-of-Sample Performance}\label{OutS.Per}
While focusing on portfolio optimization, we need to evaluate the performance of the corresponding portfolio strategies in an out-of-sample setting. For this, we apply the forecasting models and vine copula estimation to our dataset and use the rolling window method to simulate one-step-ahead returns and compute one-step-ahead asset weights.\footnote{See \cite{sahamkhadam2018portfolio} for more information on the steps required for rolling window estimation.} We use a fixed rolling window with 500 observations for each stock market. This technique can create a setting where the investor uses the available information set, forecasts tomorrow's returns, and optimizes the portfolio. Then, tomorrow, the investor obtains profit or loss from the portfolio strategy. The difficulty for the investor is twofold. First, the investor can use all the information he/she has access to (an extending window) or only the last 500 observation. However, one has to note that vine copula estimation is a time-exhausting process and including all the available data might decelerate the pace of recovery from the crisis. Second, the investor has to bear the transaction costs (TCs) if the intension is to re-balance the portfolio every day. In this case, we use the proportional TCs at 10 basis points.

Table \ref{OutS.SR} reports the SR portfolio back-testing results. From the results for both sample periods, simple Clayton, Frank, Joe and Gumbel copula models show similar performance in maximizing the investor's utility function. This indicates that during the financial crisis, vine copula models cannot outperform simple copulas when the portfolio optimization is based on the SR strategy. The only case that vine copulas (Dvine and Cvine) give better out-of-sample SR is Student$-t$ copula. By comparing the vine copula structures for single copula families (e.g., Clayton, Frank, Joe and Gumbel), we find that the C-vine structure leads to higher SR maximization than the D-vine and R-vine structures. All the SR portfolios based on the forecasting models provide better portfolio accumulation than the two benchmarks [equally-weighted (EQW) and historical portfolios] for long-term investment.

Table \ref{OutS.CVaR} presents the results for the CVaR portfolios. From a comparison of the different copula models based on the portfolios' out-of-sample CVaR, which is the optimization objective, we find that Frank and Student$-t$ vine copulas perform better than Clayton and Joe copulas. This indicates that even Clayton copulas can overestimate the downside risk during a financial crisis. In general, all the CVaR portfolio strategies obtained with the vine copula models outperform the EQW and historical portfolios in minimizing CVaR during the extended out-of-sample period (Panel B). In addition, for the Clayton, Joe, and Frank copulas, the vine models result in lower CVaR compared to the simple multivariate copulas in the same families, indicating better results from vine copula modeling during financial crisis. In most cases, Dvine and Cvine models outperform Rvine interms of economic performance. Considering the portfolio accumulation wealth, a naive EQW portfolio leads to a higher terminal value even without considering the TCs.

The MV portfolio back-testing results are shown in Table \ref{OutS.MV}. None of the forecasting models can reduce portfolio volatility better than historical MV optimization.

\subsection{Short-term Investment}
\begin{figure}[H]
	\centering
	\caption{Realized Sharpe Ratio for SR Portfolio Strategies}\label{Fig5}
	\includegraphics[width=0.9\textwidth]{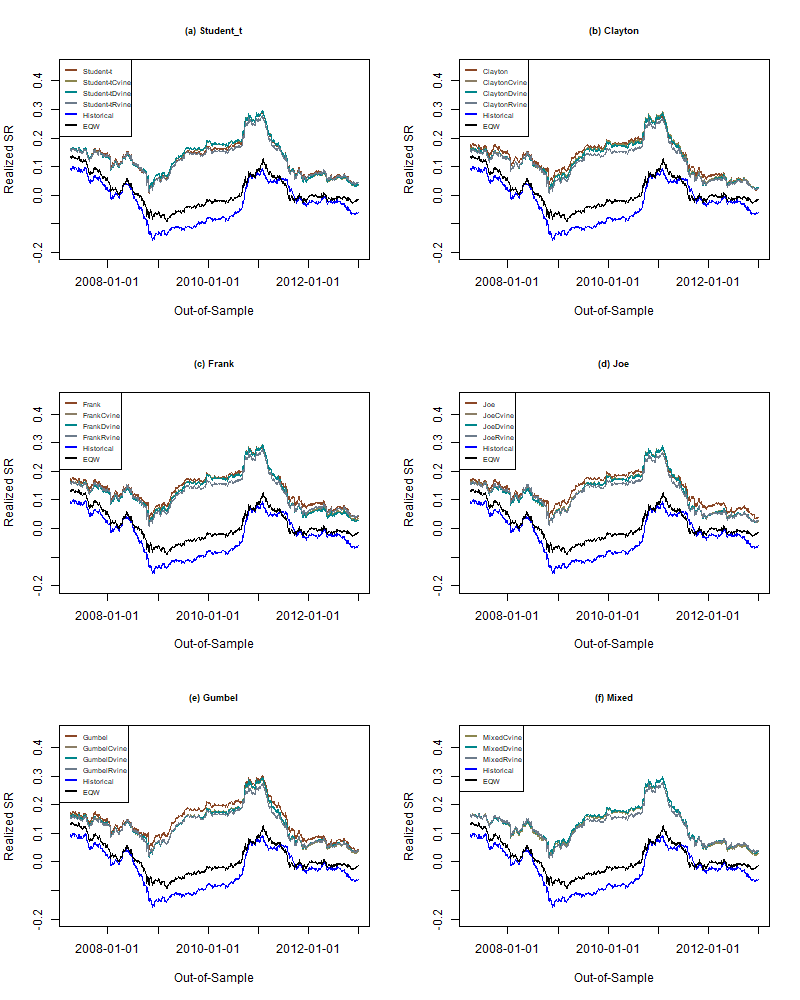}
	\caption*{Notes: This figure provides the rolling Sharpe ratio for SR portfolio strategies with a 500-day investment horizon.}
\end{figure}

\begin{figure}[H]
	\centering
	\caption{Realized CVaR for CVaR Portfolio Strategies}\label{Fig6}
	\includegraphics[width=0.9\textwidth]{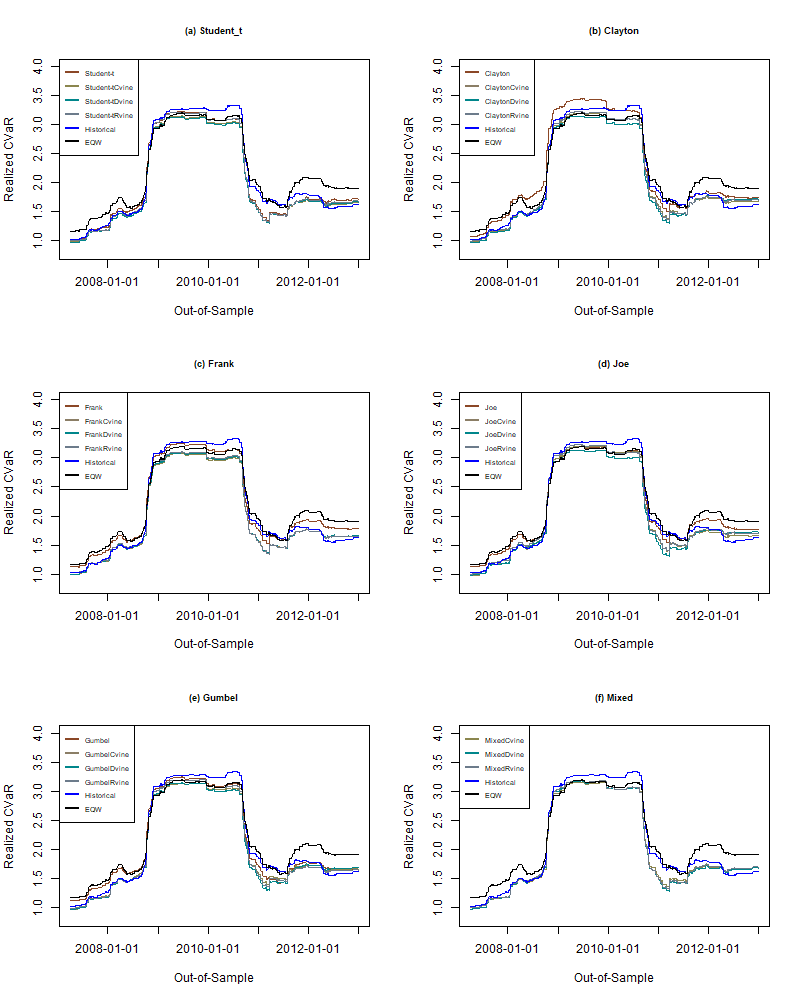}
	\caption*{Notes: This figure provides rolling CVaR for CVaR portfolio strategies with a 500-day investment horizon.}
\end{figure}

\begin{figure}[H]
	\centering
	\caption{Realized Standard Deviation for GMV Portfolio Strategies}\label{Fig7}
	\includegraphics[width=0.9\textwidth]{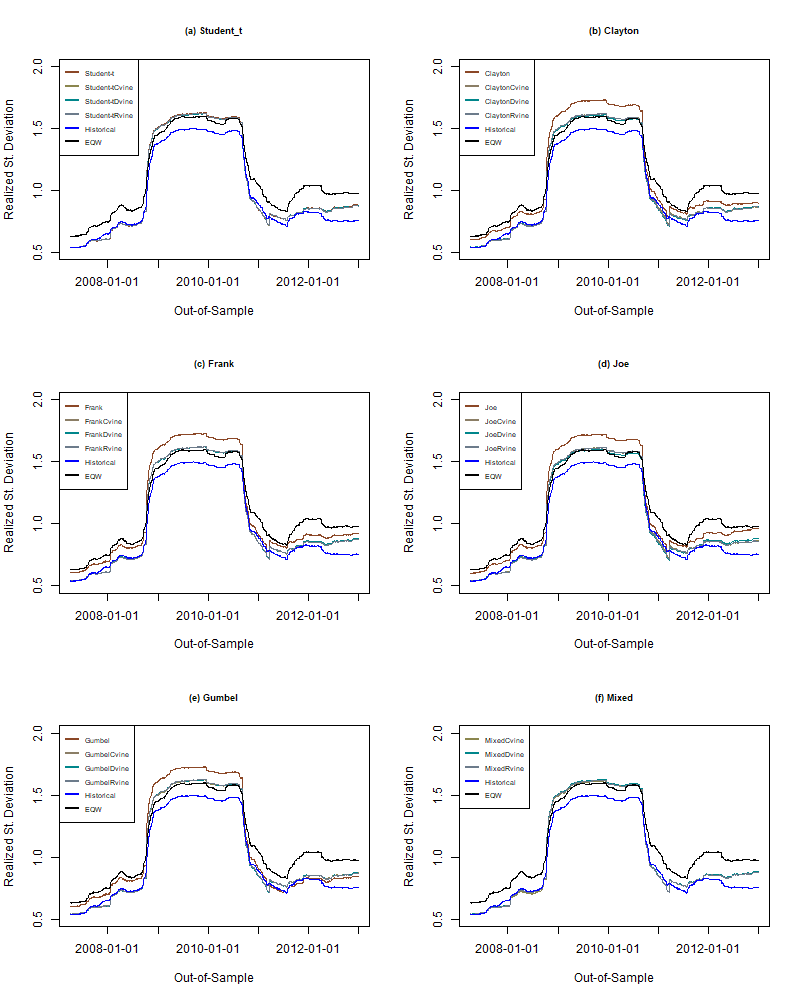}
	\caption*{Notes: This figure provides rolling standard deviation for GMV portfolio strategies with a 500-day investment horizon.}
\end{figure}

The empirical results discussed in Section \ref{OutS.Per} are based on long-term investment. However, a comparison of the competing models based on a shorter-horizon investment can show the gain from using forecasting models, particularly based on vine copula modeling. For this, we consider a two-year investment period and calculate the portfolio’s performance measures including SR, CVaR and volatility. Then, we roll this holding period and calculate the realized measures at the end of each investment period. This technique results in deeper insight into the performance of each portfolio strategy during the financial crisis. Figures \ref{Fig5}-\ref{Fig7} illustrate the results, including rolling realized SR (CVaR/volatility) for SR (CVaR/GMV) portfolios. As shown, for SR portfolios, all the forecasting models lead to higher SR, regardless of starting point of the two-year investment. Furthermore, during the financial crisis, there is slight gain from vine copula models (e.g. Dvine) in reducing the downside risk for CVaR portfolios. As regards the GMV portfolio strategies, the historical portfolio results in lower volatility for holding periods ending during the global financial crisis.

\subsection{VaR Back-testing}\label{VaR}
As our final empirical investigation, we compare the VaR back-testing results for portfolio strategies based on different copula modeling. We forecast a 1\% VaR based on simulated one-step-ahead returns for each forecasting model and compare the results with the corresponding portfolio out-of-sample returns. We conduct three tests, the unconditional coverage (UC) test proposed by \cite{kupiec1995techniques}, the conditional coverage (CC) test proposed by \cite{christoffersen1998evaluating} and the dynamic quantile (DQ) test suggested by \cite{engle2004caviar}. The UC test examines whether the actual VaR exceedance frequency is consistent with the expected one. The CC test also assumes independent number of exceedances. Since the null hypothesis of the UC test is correct exceedances, a portfolio strategy with acceptable VaR forecasts needs to reduce the test statistics until the test is insignificant \citep{kupiec1995techniques}. For the CC test, the null hypothesis is both correct exceedances and independence of VaR violations\citep{christoffersen1998evaluating}. As regards to DQ test, based on a linear regression method, the null hypothesis is that the violation are uncorrelated \citep{engle2004caviar}.

Table \ref{VaRB} presents the VaR back-testing results at different significance levels for each copula-based portfolio strategy. In general, during the global financial crisis, all of the forecasting models fail to provide adequate VaR forecasts for SR portfolio strategies. However, for GMV and CVaR strategies, in some cases, the VaR forecasts pass UC and CC tests at 5\% significance level (e.g. Clayton Cvine and Dvine). Moreover, during the crisis period Clayton copulas lead to lower number of exceedances(NE), mean absolute deviation between the observations and the quantile (AD), and average quantile loss (AQL). On the other hand, over the full out-of-sample period, Gumbel, Student$-t$ and mixed vine copulas show better results in passing the UC, CC and DQ tests and the number of exceedances.

\subsection{ES Back-testing}\label{VaR}
Due to the fact that the VaR tests in previous section have low power, we also perform ES back-testing. In doing so, we use exceedance residuals (ER) with the null hypothesis of $i.i.d.$ residuals for violations, and for VaR and ES, we use the conditional calibration (Cond. C)testing of the null hypothesis of calibrated sequence of forecasts \citep[see][]{mcneil2000estimation,nolde2017elicitability}. We also use Expected Shortfall Regression (ESR) test proposed by \cite{bayer2018regression}, where the null hypothesis is correctly specified ES forecasts.

Table \ref{ESB} reports the ES back-testing results at different significance levels for the crisis period. As the table shows, none of the models can pass all the tests. However, simple Clayton can pass the Cond. C, ER and ESR intercept tests. For GMV portfolios, Clayton and Joe show better results in accepting the null hypothesis of Cond. C and ER tests. On the other hand, Table \ref{ESB1} provides the results for whole out-of-sample period, where the Frank copula family fails to provide acceptable forecasts for CVaR; this agrees with the VaR back-testing results. The Clayton family in all copula structures and Joe vine copulas increase the test statistic for both the conditional calibration and exceedance residuals tests and accept the null hypotheses of $i.i.d.$ residuals and calibrated sequence of forecasts. Moreover, Student$-t$ copula models cannot pass all of the tests for CVaR portfolio strategy. As regards the SR portfolio, the Student$-t$ C-vine and D-vine models do not pass the exceedance residuals test, but provide acceptable results for the MV strategy. Joe, Gumbel and mixed versions of the vine copula models accept the null hypotheses for both the SR and MV portfolios. However, for the CVaR portfolio, they fail the test in a few cases. In general, these results suggest better CVaR forecasts from copula families that can model asymmetric tail dependence (e.g. Clayton).

\subsection{Regression Results}\label{Regression}
To establish more systematic evidence of the effects of the various out-of-sample copula-based portfolio strategies presented in Section \ref{OutS.Per}, we perform regressions using the various target measure outcomes of each copula model as the dependent variable. That is, for the max SR optimization strategy we define SR, for the min CVaR strategy we define CVaR, and for the GMV strategy we define the standard deviaton (StdDev) of portfolio returns as the dependent variable. Using this strategy gives the outcomes of 25 copula-based portfolio strategies at a quarterly level from Q2/2005 to Q4/2012, resulting in a total of 775 observations. In panel B we analyze the subperiod Q1/2007 to Q4/2009 for the 25 strategies, giving us 300 observations. Regression results are presented in Table \ref{Reg}, where each copula-based portfolio strategy is defined as the categorical variable. The equally weighted portfolio constitutes the reference category. The estimated coefficients of our regressions represent the differences between the copula-based portfolio strategies and the EQW portfolio.

 The results highlight that all copula-based portfolio strategies have statistically significant higher SRs when compared to the EQW strategy. Most copula-based strategies also lead to lower downside risk and lower volatility compared to the EQW portfolio. All vine copula-based strategies are also better than the optimal portfolios determined from historical returns for SR, and mostly also for downside risk measured as CVaR. Another pattern that emerges is that for SR optimal portfolios, simple copula models appear to perform better than the vines. We find that the simple Gumbel copula-based model has a SR that is on average 0.101 higher than the EQW portfolio, followed by the simple Joe copula model that has a SR of 0.097 higher than the SR of the EQW portfolio. Comparing the different copula families, it appears that Student-$t$ vines are best in terms of SR compared to the other copula families including the mixed vines. There is also a tendency that the D-vine models show higher SR compared to the R- and C-vines of the same copula family. In terms of downside risk reduction measured by CVaR, we find that the Student-$t$ based D-vine copula model leads to the greatest reduction of CVaR with 0.162 compared to EQW portfolio.

 In general, when we compare panels A and B specifically for the financial crisis, the effects from the copula-based models become more pronounced.  That is, relative to the EQW portfolio, copula-based models lead to higher SR, lower CVaR and lower volatility during the financial crisis. Again, D-vine copula models have higher SR and lower CVaR compared to the other vines. Also, comparing the various copula families it appears that Student-$t$ based vines show the greatest effects in terms of improving SR but also in reducing CVaR during the financial crisis. Surprisingly, despite their flexibility, mixed vines do not show superior performance compared to the single family models.

\section{Conclusions}\label{Conclusion}
In this study, we analyze the symmetric and asymmetric properties of international stock markets based on vine copula modeling in the 2008--2009 financial crisis portfolio management setting. Our focus is on the performance of vine copula modeling in portfolio optimization based on three utility functions, maximum SR, MV, and Min-CVaR. We examine the R-vine, C-vine, and D-vine copula structures. To account for the symmetry and asymmetry of returns during the financial crisis, we suggest four copula modeling versions: the (1) Clayton (asymmetric lower tail dependence) version with all rotations; (2) Joe (asymmetric upper tail dependence) version with all rotations; (3) Frank (symmetric no tail dependence) version; and (4) mixed version. The mixed version includes the Gaussian (no tail dependence); Student$-t$ (symmetric upper and lower tail dependence); Clayton (with all rotations); Frank; Gumbel (asymmetric upper tail dependence with all rotations); Joe (with all rotations); BB1 and BB7 (asymmetric lower and upper tail dependence with all rotations); and BB6 and BB8 (asymmetric upper tail dependence with all rotations) copulas.

The econometric analyses of portfolio out-of-sample outcomes highlight that all copula-based portfolio strategies lead to statistically significant higher SRs when compared to the EQW strategy. Copula-based strategies have statistically significant lower downside risk and lower volatility compared to the EQW portfolio. Specifically for the financial crisis, the effects from using the copula-based models become even more pronounced. That is, relative to the EQW portfolio, copula-based models lead to statistically significant higher SR, lower CVaR and lower volatility during the financial crisis.

In terms of downside risk reduction measured by CVaR, we find that the Student-$t$ based D-vine copula model leads to the greatest reduction of CVaR compared to the other vines, both for the entire period 2005--2012 and during the financial crisis 2008--2009. Thus, vines with asymmetric properties do in general not lead to better out-of-sample risk performance compared to the symmetric Student-$t$ vines. Surprisingly, mixed vines do not show superior risk performance compared to the single family vine models, despite their flexibility.

The results of this study have important implications for portfolio management during a financial crisis. Our study provides novel insights on the effects of vine copula modeling in portfolio allocation techniques. Although our results are based on the targeted dataset (including daily returns of stock markets) of the 2008--2009 financial crisis and the~pre- and post-crisis periods, this analysis can be extended to other periods as well. Studying the effects from using different return frequencies for the portfolio optimization using vines (e.g.,~weekly or monthly or even intra-day) is left for future research.


\bibliography{mybibfile}

\begin{landscape}
	\appendix
\section{Pair-copula Construction}\label{A}
\begin{table}[H]
	\centering
		\resizebox{1.3\textheight}{!}{
		\begin{threeparttable}
			\caption{Copula Families and Tail Dependence}\label{Cop.Fam}
			\begin{tabular}{lllll}
				\toprule
				\multirow{2}[4]{*}{\textbf{Copula Family}} & \multicolumn{1}{c}{Copula Function} & \multicolumn{1}{c}{Generator Function} &  \multicolumn{1}{c}{Lower Tail} & \multicolumn{1}{c}{Upper Tail} \\
				\cmidrule{2-5}
				&  $C(u_1,u_2)$    &  $\varphi(u)$     &   $\lambda_L$     & $\lambda_U$ \\
				\midrule
				1. Gaussian & $\Phi_\rho(\Phi^{-1}(u_1),\Phi^{-1}(u_2))$    & -   & 0     & 0 \\
				2. Student$-t$ & $t_{\rho,\nu}(t_\nu^{-1}(u_1),t_\nu^{-1}(u_2))$    & -    & $2t_{\nu+1}(-\sqrt{\nu+1}\sqrt{\frac{1-\rho}{1+\rho}})$     & $2t_{\nu+1}(-\sqrt{\nu+1}\sqrt{\frac{1-\rho}{1+\rho}})$ \\
				3. Clayton & $\varphi^{-1}(\varphi(u_1),\varphi(u_2))$    & $1/\theta(u^{-\theta}-1)$  & $2^{-1/\theta}$     & 0 \\
				4. Gumbel & $\varphi^{-1}(\varphi(u_1),\varphi(u_2))$    & $(-\text{log}u)^\theta$   &  0     & $2-2^{1/\theta}$ \\
				5. Frank & $\varphi^{-1}(\varphi(u_1),\varphi(u_2))$    & $-\text{log}[(e^{-\theta u}-1)/(e^{-\theta}-1)]$  & 0     & 0 \\
				6. Joe & $\varphi^{-1}(\varphi(u_1),\varphi(u_2))$    & $-\text{log}[1-(1-u)^\theta]$  & 0     & $2-2^{1/\theta}$ \\
				7. BB1 & $\varphi^{-1}(\varphi(u_1),\varphi(u_2))$    & $(u^{-\theta}-1)^\delta$   & $2^{-1/(\theta\delta)}$     & $2-2^{1/\delta}$ \\
				8. BB6 & $\varphi^{-1}(\varphi(u_1),\varphi(u_2))$    &  $(-\text{log}[1-(1-u)^\theta])^\delta$  & 0 & $2-2^{1/(\delta\theta)}$ \\
				9. BB7 & $\varphi^{-1}(\varphi(u_1),\varphi(u_2))$    & $(1-(1-u)^\theta)^{-\delta}-1$    & $2^{-1/\delta}$ & $2-2^{1/\theta}$ \\
				10. BB8 & $\varphi^{-1}(\varphi(u_1),\varphi(u_2))$    & $-\text{log}[(1-(1-\delta u)^\theta)/(1-(1-\delta)^\theta)]$    & 0 & $2-2^{1/\theta}$ if $\delta=1$ otherwise $0$ \\
				13. rotated Clayton ($180^o$) & $u_1+u_2-1+C(1-u_1,1-u_2)$    & $1/\theta(u^{-\theta}-1)$   & 0 & $2^{-1/\theta}$ \\
				14. rotated Gumbel ($180^o$) & $u_1+u_2-1+C(1-u_1,1-u_2)$    & $(-\text{log}u)^\theta$    & $2-2^{1/\theta}$ & 0 \\
				16. rotated Joe ($180^o$) & $u_1+u_2-1+C(1-u_1,1-u_2)$    & $-\text{log}[1-(1-u)^\theta]$   & $2-2^{1/\theta}$ & 0 \\
				17. rotated BB1 ($180^o$) & $u_1+u_2-1+C(1-u_1,1-u_2)$    & $(u^{-\theta}-1)^\delta$   & $2-2^{1/\delta}$ & $2^{-1/(\delta\theta)}$ \\
				18. rotated BB6 ($180^o$) & $u_1+u_2-1+C(1-u_1,1-u_2)$    & $(-\text{log}[1-(1-u)^\theta])^\delta$    & $2-2^{1/(\delta\theta)}$ & 0 \\
				19. rotated BB7 ($180^o$) & $u_1+u_2-1+C(1-u_1,1-u_2)$    & $(1-(1-u)^\theta)^{-\delta}-1$   & $2-2^{1/\theta}$ & $2^{-1/\delta}$ \\
				20. rotated BB8 ($180^o$) & $u_1+u_2-1+C(1-u_1,1-u_2)$    & $-\text{log}[(1-(1-\delta u)^\theta)/(1-(1-\delta)^\theta)]$    & $2-2^{1/\theta}$ if $\delta=1$ otherwise $0$ & 0 \\
				23. rotated Clayton ($90^o$) & $u_2-C(1-u_1,u_2)$    & $1/\theta(u^{-\theta}-1)$    & 0 & 0 \\
				24. rotated Gumbel ($90^o$) & $u_2-C(1-u_1,u_2)$    & $(-\text{log}u)^\theta$    & 0 & 0 \\
				26. rotated Joe ($90^o$) & $u_2-C(1-u_1,u_2)$    & $-\text{log}[1-(1-u)^\theta]$    & 0 & 0 \\
				27. rotated BB1 ($90^o$) & $u_2-C(1-u_1,u_2)$    & $(u^{-\theta}-1)^\delta$    & 0 & 0 \\
				28. rotated BB6 ($90^o$) & $u_2-C(1-u_1,u_2)$    & $(-\text{log}[1-(1-u)^\theta])^\delta$    & 0 & 0 \\
				29. rotated BB7 ($90^o$) & $u_2-C(1-u_1,u_2)$    & $(1-(1-u)^\theta)^{-\delta}-1$    & 0 & 0 \\
				30. rotated BB8 ($90^o$) & $u_2-C(1-u_1,u_2)$    & $-\text{log}[(1-(1-\delta u)^\theta)/(1-(1-\delta)^\theta)]$    & 0 & 0 \\
				33. rotated Clayton ($270^o$) & $u_1-C(u_1,1-u_2)$    & $1/\theta(u^{-\theta}-1)$   & 0 & 0 \\
				34. rotated Gumbel ($270^o$) & $u_1-C(u_1,1-u_2)$    & $(-\text{log}u)^\theta$    & 0 & 0 \\
				36. rotated Joe ($270^o$) & $u_1-C(u_1,1-u_2)$    & $-\text{log}[1-(1-u)^\theta]$   & 0 & 0 \\
				37. rotated BB1 ($270^o$) & $u_1-C(u_1,1-u_2)$    & $(u^{-\theta}-1)^\delta$   & 0 & 0 \\
				38. rotated BB6 ($270^o$) & $u_1-C(u_1,1-u_2)$    & $(-\text{log}[1-(1-u)^\theta])^\delta$    & 0 & 0 \\
				39. rotated BB7 ($270^o$) & $u_1-C(u_1,1-u_2)$    & $(1-(1-u)^\theta)^{-\delta}-1$    & 0 & 0 \\
				40. rotated BB8 ($270^o$) & $u_1-C(u_1,1-u_2)$    & $-\text{log}[(1-(1-\delta u)^\theta)/(1-(1-\delta)^\theta)]$    & 0 & 0 \\
				\bottomrule
			\end{tabular}%
			\begin{tablenotes}
				\small
				\item Note: This table presents the copula and generator functions, the Kendall $\tau$, and lower and upper tail dependence of different bivariate copula families. For more details of the parameter restrictions and Kendall's $\tau$, see \cite{brechmann2013modeling}.
			\end{tablenotes}
	\end{threeparttable}}
\end{table}
\end{landscape}

\section{Financial Data}\label{B}
\begin{table}[H]
	\small
	\centering
	\begin{threeparttable}
		\caption{ Financial Data Descriptive Statistics}\label{Desc.Stat}
		\begin{tabular}{lcccccccccc}
			\toprule
			Series & Mean  & St. Deviation & Median & Minimum & Maximum & Skewness & Kurtosis & VaR   & CVaR  & JB \\
			\midrule
			\multicolumn{11}{l}{\textit{Panel A: 2008-2009}} \\
			S\&P 500 & -0.05 & 2.16  & 0.03  & -9.47 & 10.96 & -0.12 & 4.53  & 2.41  & 4.19  & 454*** \\
			FTSE 100 & -0.03 & 1.94  & 0.00     & -9.27 & 9.38  & -0.01 & 4.42  & 2.28  & 3.65  & 432*** \\
			DAX 30 & -0.06 & 2.08  & 0.00    & -7.43 & 10.8  & 0.31  & 4.69  & 2.19  & 3.91  & 493*** \\
			IBEX 35 & -0.05 & 2.07  & 0.00     & -9.59 & 10.12 & 0.04  & 3.92  & 2.33  & 3.88  & 339*** \\
			KOSPI & -0.02 & 2.02  & 0.01  & -11.17 & 11.28 & -0.47 & 5.67  & 2.09  & 3.93  & 729*** \\
			TOPIX & -0.09 & 2.12  & 0.00     & -10.01 & 12.86 & -0.1  & 4.72  & 2.42  & 4.05  & 493*** \\
			S\&P/TSX & -0.03 & 2.07  & 0.09  & -9.79 & 9.37  & -0.49 & 3.7   & 2.49  & 4.13  & 322*** \\
			OMXS 30 & -0.02 & 2.18  & 0.00     & -7.51 & 9.87  & 0.29  & 2.28  & 2.51  & 3.95  & 122*** \\
			SMI   & -0.05 & 1.77  & 0.00     & -8.11 & 10.79 & 0.23  & 4.82  & 1.92  & 3.29  & 518*** \\
			OMXH  & -0.11 & 2.11  & -0.05 & -7.92 & 8.85  & 0.19  & 1.97  & 2.53  & 3.91  & 89*** \\
			MERVAL & 0.01  & 2.51  & 0.01  & -12.95 & 10.43 & -0.59 & 4.07  & 2.88  & 5.06  & 396*** \\
			HERMES & -0.09 & 2.14  & 0.00     & -17.2 & 5.76  & -1.39 & 8.67  & 2.60  & 4.32  & 1826*** \\
			\multicolumn{11}{l}{\textit{Panel B: 2003-2012}} \\
			S\&P 500 & 0.02  & 1.29  & 0.05  & -9.47 & 10.96 & -0.31 & 11.07 & 1.25  & 2.39  & 12842*** \\
			FTSE 100 & 0.01  & 1.21  & 0.01  & -9.27 & 9.38  & -0.15 & 8.88  & 1.22  & 2.23  & 8236*** \\
			DAX 30 & 0.04  & 1.41  & 0.07  & -7.43 & 10.8  & 0.04  & 6.65  & 1.54  & 2.59  & 4623*** \\
			IBEX 35 & 0.01  & 1.49  & 0.05  & -9.59 & 13.48 & 0.14  & 7.62  & 1.56  & 2.79  & 6077*** \\
			KOSPI & 0.04  & 1.44  & 0.05  & -11.17 & 11.28 & -0.57 & 6.54  & 1.59  & 2.75  & 4606*** \\
			TOPIX & 0.00     & 1.39  & 0.00     & -10.01 & 12.86 & -0.45 & 8.79  & 1.52  & 2.59  & 8145*** \\
			S\&P/TSX & 0.02  & 1.19  & 0.05  & -9.79 & 9.37  & -0.72 & 10.78 & 1.18  & 2.26  & 12351*** \\
			OMXS 30 & 0.03  & 1.46  & 0.03  & -7.51 & 9.87  & 0.03  & 4.59  & 1.56  & 2.72  & 2201*** \\
			SMI   & 0.01  & 1.13  & 0.04  & -8.11 & 10.79 & -0.02 & 8.59  & 1.23  & 2.10  & 7698*** \\
			OMXH  & 0     & 1.48  & 0.02  & -9.23 & 8.85  & -0.19 & 4.48  & 1.69  & 2.81  & 2110*** \\
			MERVAL & 0.06  & 1.86  & 0.04  & -12.95 & 10.43 & -0.58 & 5.07  & 2.02  & 3.50  & 2820*** \\
			HERMES & 0.08  & 1.68  & 0.04  & -17.2 & 13.7  & -0.9  & 10.94 & 1.72  & 3.18  & 12838*** \\
			\bottomrule
		\end{tabular}%
		\begin{tablenotes}
			\small
			\item Note: This Table provides the descriptive statistics for daily  returns of 12 stock indexes. The returns are in logarithmic form. In Panel A, the sample period is from January 1, 2008 to December 31, 2009. Panel B reports the descriptive statistics from June 3, 2003 to December 31, 2012. The VaR and CVaR are calculated empirically at 10\% level. JB shows the Jarque-Bera normality test results. *** denotes significance at the 1\% level.
		\end{tablenotes}
	\end{threeparttable}
\end{table}
\begin{landscape}
\section{Empirical Analysis}\label{C}
\begin{table}[H]
	\centering
		\begin{threeparttable}
			\caption{AR-GARCH Parameter Estimation}\label{AR.GARCH.Param}
			\begin{tabular}{lcccccccccccc}
				\toprule
				Parameter & S\&P 500 & FTSE 100 & DAX 30 & IBEX 35 & KOSPI & TOPIX & S\&P/TSX & OMXS 30 & SMI   & OMXH  & MERVAL & HERMES \\
				\midrule
				\multicolumn{13}{l}{\textit{Panel A: 2008-2009}} \\
				$\mu_j$    & 0.014 & 0.044 & -0.001 & 0.022 & 0.018 & -0.065 & 0.029 & 0.025 & 0.032 & -0.052 & 0.053 & 0.014 \\
				$\phi_j$   & -0.115*** & -0.092** & -0.062* & -0.066 & -0.033 & -0.074* & -0.079** & -0.079** & -0.039 & -0.049 & -0.022 & 0.172*** \\
				$\omega_j$ & 0.022 & 0.054* & 0.063 & 0.069 & 0.040 & 0.087 & 0.027 & 0.024 & 0.041 & 0.018 & 0.092 & 0.371 \\
				$\alpha_{1j}$ & 0.092*** & 0.083*** & 0.078** & 0.092*** & 0.073** & 0.123*** & 0.099*** & 0.064*** & 0.104*** & 0.055*** & 0.096** & 0.23* \\
				$\beta_{1j}$ & 0.903*** & 0.899*** & 0.907*** & 0.89*** & 0.917*** & 0.854*** & 0.894*** & 0.93*** & 0.88*** & 0.94*** & 0.897*** & 0.705*** \\
				$\zeta_j$    & 0.887*** & 0.957*** & 0.964*** & 0.942*** & 0.883*** & 0.904*** & 0.755*** & 1.029*** & 0.97*** & 1.024*** & 0.913*** & 0.908*** \\
				$\eta_j$    & 8.294*** & 7.348*** & 6.249*** & 10.618* & 5.302*** & 12.932* & 13.336** & 27.717 & 10.552** & 7.504*** & 4.263*** & 4.675*** \\
				\multicolumn{13}{l}{\textit{Panel B: 2003-2012}} \\
				$\mu_j$    & 0.047*** & 0.046*** & 0.079*** & 0.064*** & 0.086*** & 0.047*** & 0.058*** & 0.070*** & 0.048*** & 0.064*** & 0.084*** & 0.147*** \\
				$\phi_j$   & -0.068*** & -0.067*** & -0.036* & -0.01 & -0.028* & 0.01  & -0.01 & -0.066*** & -0.02 & 0.02  & 0.02  & 0.115*** \\
				$\omega_j$ & 0.011*** & 0.011*** & 0.017*** & 0.012*** & 0.027*** & 0.029** & 0.012*** & 0.015** & 0.018*** & 0.013** & 0.111*** & 0.209*** \\
				$\alpha_{1j}$ & 0.080*** & 0.091*** & 0.083*** & 0.087*** & 0.074*** & 0.090*** & 0.083*** & 0.076*** & 0.100*** & 0.058*** & 0.093*** & 0.204*** \\
				$\beta_{1j}$ & 0.913*** & 0.901*** & 0.909*** & 0.910*** & 0.912*** & 0.893*** & 0.904*** & 0.918*** & 0.882*** & 0.937*** & 0.878*** & 0.746*** \\
				$\zeta_j$    & 0.911*** & 0.902*** & 0.907*** & 0.931*** & 0.887*** & 0.918*** & 0.846*** & 0.912*** & 0.900*** & 0.948*** & 0.933*** & 0.971*** \\
				$\eta_j$    & 6.049*** & 10.000*** & 7.719*** & 6.850*** & 6.710*** & 8.911*** & 9.562*** & 7.650*** & 8.865*** & 5.899*** & 4.626*** & 3.967*** \\
				\bottomrule
			\end{tabular}%
			\begin{tablenotes}
				\footnotesize
				\item Note: This table presents an in-sample AR-GARCH model estimation for 12 stock index daily returns. In Panel A, the sample period is from January 1, 2008 to December 31, 2009. Panel B reports the results from June 3, 2003 to December 31, 2012. $\mu_j$ is a constant in the mean equation, $\phi_j$ is the AR(1) coefficient, $\omega_j$ is a constant in the volatility equation, $\alpha_{1j}$ and $\beta_{1j}$ are GARCH(1,1) coefficients, and $\zeta_j$ and $\eta_j$ give the skewness and shape parameters, respectively. ***, **, and * denote significance at the 1\%, 5\%, and 10\% levels, respectively.
			\end{tablenotes}
	\end{threeparttable}
\end{table}
\end{landscape}

\begin{table}[H]
	\resizebox{\textwidth}{!}{
		\begin{threeparttable}
			\caption{Goodness-of-fit Testing for Copula Models}\label{GOF}
			\begin{tabular}{llllllllllllll}
				\toprule
				\multicolumn{1}{l}{\multirow{2}[4]{*}{Sample Period}} & \multicolumn{1}{l}{\multirow{2}[4]{*}{Vine Structure}} & \multicolumn{2}{l}{\multirow{2}[4]{*}{Test}} & \multicolumn{2}{l}{Student$-t$} & \multicolumn{2}{l}{Clayton} & \multicolumn{2}{l}{Frank} & \multicolumn{2}{l}{Joe} & \multicolumn{2}{l}{Gumbel} \\
				\cmidrule{5-14}          &       & \multicolumn{2}{l}{} & $t-$statistic & $p-$value & $t-$statistic & $p-$value & $t-$statistic & $p-$value & $t-$statistic & $p-$value & $t-$statistic & $p-$value \\
				\midrule
				\textit{Panel A: 2008-2009} & Cvine & ECP   & CvM   & 3.90  & 0.53  & 3.74  & 0.64  & 3.84  & 0.61  & 4.13  & 0.54  & 3.92  & 0.54 \\
				&       &       & KS    & 6.99  & 0.55  & 7.08  & 0.43  & 6.59  & 0.74  & 7.22  & 0.35  & 7.13  & 0.43 \\
				&       & ECP2  & CvM   & 0.00  & 1.00  & 0.00  & 1.00  & 0.00  & 1.00  & 0.00  & 1.00  & 0.00  & 1.00 \\
				&       &       & KS    & 0.18  & 1.00  & 0.23  & 0.99  & 0.17  & 1.00  & 0.41  & 0.53  & 0.20  & 1.00 \\
				& Dvine & ECP   & CvM   & 3.48  & 0.85  & 3.92  & 0.69  & 3.77  & 0.64  & 3.98  & 0.62  & 3.95  & 0.60 \\
				&       &       & KS    & 6.94  & 0.57  & 6.79  & 0.70  & 6.5   & 0.84  & 6.92  & 0.51  & 7.03  & 0.32 \\
				&       & ECP2  & CvM   & 0.00  & 1.00  & 0.00  & 1.00  & 0.00  & 1.00  & 0.01  & 1.00  & 0.00  & 1.00 \\
				&       &       & KS    & 0.22  & 0.94  & 0.51  & 0.43  & 0.77  & 0.36  & 0.81  & 0.34  & 0.20  & 0.99 \\
				& Rvine & ECP   & CvM   & 4.00  & 0.56  & 4.10  & 0.48  & 4.01  & 0.54  & 4.26  & 0.37  & 4.21  & 0.42 \\
				&       &       & KS    & 6.66  & 0.67  & 7.15  & 0.36  & 6.77  & 0.55  & 6.79  & 0.57  & 6.82  & 0.63 \\
				&       & ECP2  & CvM   & 0.00  & 1.00  & 0.00  & 1.00  & 0.00  & 1.00  & 0.00  & 1.00  & 0.00  & 1.00 \\
				&       &       & KS    & 0.25  & 0.7   & 0.42  & 0.48  & 0.31  & 0.88  & 0.37  & 0.64  & 0.14  & 1.00 \\
				\textit{Panel B: 2003-2012} & Cvine & ECP   & CvM   & 8.74  & 0.66  & 9.19  & 0.45  & 8.85  & 0.61  & 8.72  & 0.64  & 9.08  & 0.51 \\
				&       &       & KS    & 13.08 & 0.67  & 14.08 & 0.28  & 14.38 & 0.21  & 13.63 & 0.44  & 14.78 & 0.06 \\
				&       & ECP2  & CvM   & 0.00  & 1.00  & 0.01  & 0.93  & 0.00  & 1.00  & 0.01  & 0.94  & 0.00  & 1.00 \\
				&       &       & KS    & 0.34  & 0.66  & 1.63  & 0.37  & 0.52  & 0.61  & 1.74  & 0.43  & 0.65  & 0.36 \\
				& Dvine & ECP   & CvM   & 8.94  & 0.58  & 9.06  & 0.45  & 8.79  & 0.66  & 9.16  & 0.38  & 8.50  & 0.78 \\
				&       &       & KS    & 13.78 & 0.36  & 14.93 & 0.05  & 13.48 & 0.67  & 13.83 & 0.41  & 14.43 & 0.18 \\
				&       & ECP2  & CvM   & 0.00  & 1.00  & 0.01  & 0.91  & 0.00  & 0.96  & 0.01  & 0.88  & 0.00  & 1.00 \\
				&       &       & KS    & 0.33  & 0.51  & 1.35  & 0.62  & 1.96  & 0.21  & 1.69  & 0.51  & 0.56  & 0.56 \\
				& Rvine & ECP   & CvM   & 9.01  & 0.43  & 9.53  & 0.27  & 8.11  & 0.90  & 8.93  & 0.52  & 8.58  & 0.72 \\
				&       &       & KS    & 13.43 & 0.55  & 13.53 & 0.47  & 14.08 & 0.22  & 13.93 & 0.41  & 14.53 & 0.13 \\
				&       & ECP2  & CvM   & 0.00  & 1.00  & 0.00  & 0.95  & 0.01  & 0.97  & 0.00  & 1.00  & 0.00  & 1.00 \\
				&       &       & KS    & 0.61  & 0.41  & 1.24  & 0.49  & 0.64  & 0.74  & 1.03  & 0.61  & 0.46  & 0.55 \\
				\bottomrule
			\end{tabular}%
			\begin{tablenotes}
				\footnotesize
				\item This table reports the results of copula goodness-of-fit tests. The tests include empirical copula process (ECP) and its combination with probability integral transform (ECP2). For each test, Cramer-von Mises (CvM) and Kolmogorov-Smirnov (KS) statistics are reported. The p-values are obtained using bootstrapping.
			\end{tablenotes}
	\end{threeparttable}}
\end{table}

\begin{table}[H]
	\centering
	\resizebox{\textwidth}{!}{
		\begin{threeparttable}
			\caption{In-Sample Performance of SR Portfolio Strategies}\label{InS.SR}
			\begin{tabular}{lllllllllllll}
				\toprule
				Sample Period & \multicolumn{6}{l}{\textit{Panel A: 2008-2009}} & \multicolumn{6}{l}{\textit{Panel B: 2003-2012}} \\
				\midrule
				\multicolumn{1}{r}{\multirow{2}[2]{*}{Forecasting Model}} & \multicolumn{1}{r}{\multirow{2}[2]{*}{Mean}} & \multicolumn{1}{p{4.055em}}{St.} & \multicolumn{1}{r}{\multirow{2}[2]{*}{SR}} & \multicolumn{1}{r}{\multirow{2}[2]{*}{CVaR}} & \multicolumn{1}{r}{\multirow{2}[2]{*}{STARR}} & \multicolumn{1}{p{4.055em}}{Terminal} & \multicolumn{1}{r}{\multirow{2}[2]{*}{Mean}} & \multicolumn{1}{p{4.055em}}{St.} & \multicolumn{1}{r}{\multirow{2}[2]{*}{SR}} & \multicolumn{1}{r}{\multirow{2}[2]{*}{CVaR}} & \multicolumn{1}{r}{\multirow{2}[2]{*}{STARR}} & \multicolumn{1}{p{4.055em}}{Terminal} \\
				&       & \multicolumn{1}{p{4.055em}}{Deviation} &       &       &       & \multicolumn{1}{p{4.055em}}{Wealth} &       & \multicolumn{1}{p{4.055em}}{Deviation} &       &       &       & \multicolumn{1}{p{4.055em}}{Wealth} \\
				\midrule
				EQW   & -0.050 & 1.569 & -0.032 & 3.115 & -0.016 & 72.143 & 0.027 & 1.000 & 0.027 & 1.901 & 0.014 & 172.071 \\
				Historical & -0.050 & 1.651 & -0.030 & 3.337 & -0.015 & 71.653 & 0.062 & 1.137 & 0.055 & 2.147 & 0.029 & 402.461 \\
				Student$-t$ & -0.051 & 1.480 & -0.034 & 2.964 & -0.017 & 72.369 & 0.044 & 0.961 & 0.046 & 1.819 & 0.024 & 267.195 \\
				Student$-t$ Cvine & -0.045 & 1.503 & -0.030 & 2.981 & -0.015 & 74.323 & 0.041 & 0.982 & 0.042 & 1.849 & 0.022 & 249.133 \\
				Student$-t$ Dvine & -0.045 & 1.501 & -0.030 & 2.978 & -0.015 & 74.517 & 0.042 & 0.980 & 0.042 & 1.845 & 0.023 & 250.148 \\
				Student$-t$ Rvine & -0.058 & 1.483 & -0.039 & 2.950 & -0.020 & 69.622 & 0.037 & 0.945 & 0.039 & 1.781 & 0.021 & 222.577 \\
				Clayton & -0.053 & 1.500 & -0.035 & 2.998 & -0.018 & 71.411 & 0.033 & 0.949 & 0.035 & 1.800 & 0.018 & 203.522 \\
				Clayton Cvine & -0.047 & 1.492 & -0.031 & 2.951 & -0.016 & 73.868 & 0.040 & 0.967 & 0.041 & 1.816 & 0.022 & 238.015 \\
				Clayton Dvine & -0.045 & 1.508 & -0.030 & 2.996 & -0.015 & 74.547 & 0.040 & 0.984 & 0.041 & 1.852 & 0.022 & 240.666 \\
				Clayton Rvine & -0.059 & 1.482 & -0.040 & 2.945 & -0.020 & 69.401 & 0.036 & 0.945 & 0.038 & 1.782 & 0.020 & 216.270 \\
				Frank & -0.057 & 1.496 & -0.038 & 3.007 & -0.019 & 69.900 & 0.042 & 0.963 & 0.044 & 1.829 & 0.023 & 255.151 \\
				Frank Cvine & -0.045 & 1.497 & -0.030 & 2.978 & -0.015 & 74.681 & 0.041 & 0.973 & 0.042 & 1.833 & 0.023 & 248.508 \\
				Frank Dvine & -0.045 & 1.497 & -0.030 & 2.974 & -0.015 & 74.430 & 0.041 & 0.973 & 0.042 & 1.831 & 0.023 & 248.638 \\
				Frank Rvine & -0.057 & 1.488 & -0.038 & 2.980 & -0.019 & 69.968 & 0.036 & 0.940 & 0.039 & 1.779 & 0.020 & 221.021 \\
				Joe   & -0.055 & 1.491 & -0.037 & 2.990 & -0.019 & 70.525 & 0.038 & 0.949 & 0.040 & 1.802 & 0.021 & 229.257 \\
				Joe Cvine & -0.047 & 1.490 & -0.032 & 2.944 & -0.016 & 73.694 & 0.039 & 0.964 & 0.040 & 1.811 & 0.021 & 233.745 \\
				Joe Dvine & -0.045 & 1.509 & -0.030 & 3.001 & -0.015 & 74.428 & 0.040 & 0.984 & 0.040 & 1.851 & 0.021 & 238.578 \\
				Joe Rvine & -0.057 & 1.485 & -0.038 & 2.960 & -0.019 & 70.037 & 0.035 & 0.941 & 0.037 & 1.774 & 0.019 & 211.004 \\
				Gumbel & -0.045 & 1.468 & -0.031 & 2.946 & -0.015 & 74.675 & 0.038 & 0.914 & 0.041 & 1.723 & 0.022 & 231.073 \\
				Gumbel Cvine & -0.046 & 1.500 & -0.030 & 2.974 & -0.015 & 74.223 & 0.041 & 0.979 & 0.042 & 1.841 & 0.022 & 244.966 \\
				Gumbel Dvine & -0.044 & 1.509 & -0.029 & 2.999 & -0.015 & 74.956 & 0.042 & 0.986 & 0.043 & 1.856 & 0.023 & 251.647 \\
				Gumbel Rvine & -0.058 & 1.484 & -0.039 & 2.952 & -0.020 & 69.755 & 0.036 & 0.945 & 0.038 & 1.780 & 0.020 & 217.212 \\
				Mixed Cvine & -0.046 & 1.497 & -0.031 & 2.967 & -0.016 & 73.996 & 0.041 & 0.976 & 0.042 & 1.837 & 0.022 & 245.010 \\
				Mixed Dvine & -0.045 & 1.500 & -0.030 & 2.975 & -0.015 & 74.634 & 0.042 & 0.978 & 0.043 & 1.841 & 0.023 & 250.366 \\
				Mixed Rvine & -0.057 & 1.486 & -0.038 & 2.978 & -0.019 & 70.164 & 0.036 & 0.939 & 0.039 & 1.775 & 0.021 & 221.766 \\
				\bottomrule
			\end{tabular}%
			\begin{tablenotes}
				\footnotesize
				\item Note: This table provides in-sample performance of copula-based SR portfolios strategies. Portfolio strategies are constructed by utilizing realized assets' returns and optimal weights over the targeted period. Optimal weights are obtained from simulated returns by estimating the conditional mean and volatility from AR-GARCH model in Section \ref{MarginalModeling} and estimating the joint distribution (including dependency structure) and drawing 10000 simulations from different simple and vine copulas. In mixed vines, the copula selection is based on AIC including all the families in Table \ref{Cop.Fam}. In Panel A, the sample period is from January 1, 2008 to December 31, 2009. Panel B reports the results from June 3, 2003 to December 31, 2012. All the performance measures are in percentage. STARR (mean to CVaR) ratio and CVaR are estimated empirically at 10\% level. The terminal wealth, based on a buy-and-hold strategy with \$100 investment at the beginning of each period, is also reported.
			\end{tablenotes}
	\end{threeparttable}}
\end{table}

\begin{table}[H]
	\centering
	\resizebox{\textwidth}{!}{
		\begin{threeparttable}
			\caption{In-Sample Performance of CVaR Portfolio Strategies}\label{InS.CVaR}
			\begin{tabular}{lcccccccccccc}
				\toprule
				Sample Period & \multicolumn{6}{l}{\textit{Panel A: 2008-2009}} & \multicolumn{6}{l}{\textit{Panel B: 2003-2012}} \\
				\midrule
				\multicolumn{1}{r}{\multirow{2}[2]{*}{Forecasting Model}} & \multicolumn{1}{r}{\multirow{2}[2]{*}{Mean}} & \multicolumn{1}{p{4.055em}}{St.} & \multicolumn{1}{r}{\multirow{2}[2]{*}{SR}} & \multicolumn{1}{r}{\multirow{2}[2]{*}{CVaR}} & \multicolumn{1}{r}{\multirow{2}[2]{*}{STARR}} & \multicolumn{1}{p{4.055em}}{Terminal} & \multicolumn{1}{r}{\multirow{2}[2]{*}{Mean}} & \multicolumn{1}{p{4.055em}}{St.} & \multicolumn{1}{r}{\multirow{2}[2]{*}{SR}} & \multicolumn{1}{r}{\multirow{2}[2]{*}{CVaR}} & \multicolumn{1}{r}{\multirow{2}[2]{*}{STARR}} & \multicolumn{1}{p{4.055em}}{Terminal} \\
				&       & \multicolumn{1}{p{4.055em}}{Deviation} &       &       &       & \multicolumn{1}{p{4.055em}}{Wealth} &       & \multicolumn{1}{p{4.055em}}{Deviation} &       &       &       & \multicolumn{1}{p{4.055em}}{Wealth} \\
				\midrule
				EQW   & -0.050 & 1.569 & -0.032 & 3.115 & -0.016 & 72.143 & 0.027 & 1.000 & 0.027 & 1.901 & 0.014 & 172.071 \\
				Historical & -0.061 & 1.406 & -0.043 & 2.805 & -0.022 & 69.142 & 0.031 & 0.885 & 0.035 & 1.686 & 0.018 & 196.133 \\
				Student$-t$ & -0.047 & 1.489 & -0.032 & 2.994 & -0.016 & 73.780 & 0.024 & 0.907 & 0.027 & 1.711 & 0.014 & 164.833 \\
				Student$-t$ Cvine & -0.048 & 1.456 & -0.033 & 2.924 & -0.016 & 73.554 & 0.028 & 0.903 & 0.031 & 1.703 & 0.017 & 182.531 \\
				Student$-t$ Dvine & -0.049 & 1.458 & -0.034 & 2.927 & -0.017 & 73.286 & 0.028 & 0.902 & 0.032 & 1.701 & 0.017 & 182.880 \\
				Student$-t$ Rvine & -0.050 & 1.456 & -0.035 & 2.906 & -0.017 & 72.711 & 0.027 & 0.904 & 0.030 & 1.706 & 0.016 & 175.127 \\
				Clayton & -0.054 & 1.534 & -0.035 & 3.064 & -0.018 & 70.879 & 0.024 & 0.963 & 0.025 & 1.832 & 0.013 & 162.953 \\
				Clayton Cvine & -0.052 & 1.477 & -0.035 & 2.958 & -0.018 & 71.919 & 0.025 & 0.907 & 0.028 & 1.710 & 0.015 & 168.851 \\
				Clayton Dvine & -0.050 & 1.480 & -0.034 & 2.972 & -0.017 & 72.645 & 0.026 & 0.920 & 0.028 & 1.738 & 0.015 & 171.295 \\
				Clayton Rvine & -0.050 & 1.441 & -0.035 & 2.872 & -0.018 & 72.765 & 0.028 & 0.890 & 0.031 & 1.675 & 0.017 & 180.570 \\
				Frank & -0.055 & 1.534 & -0.036 & 3.055 & -0.018 & 70.667 & 0.024 & 0.964 & 0.025 & 1.833 & 0.013 & 162.400 \\
				Frank Cvine & -0.049 & 1.468 & -0.033 & 2.956 & -0.017 & 73.194 & 0.028 & 0.907 & 0.030 & 1.717 & 0.016 & 178.928 \\
				Frank Dvine & -0.050 & 1.468 & -0.034 & 2.955 & -0.017 & 72.819 & 0.027 & 0.906 & 0.030 & 1.713 & 0.016 & 178.173 \\
				Frank Rvine & -0.050 & 1.475 & -0.034 & 2.952 & -0.017 & 72.655 & 0.026 & 0.914 & 0.028 & 1.732 & 0.015 & 169.945 \\
				Joe   & -0.053 & 1.532 & -0.035 & 3.058 & -0.017 & 71.310 & 0.025 & 0.961 & 0.026 & 1.828 & 0.014 & 165.230 \\
				Joe Cvine & -0.051 & 1.479 & -0.034 & 2.962 & -0.017 & 72.542 & 0.026 & 0.906 & 0.029 & 1.706 & 0.015 & 172.234 \\
				Joe Dvine & -0.051 & 1.487 & -0.034 & 2.979 & -0.017 & 72.166 & 0.025 & 0.924 & 0.027 & 1.744 & 0.014 & 167.082 \\
				Joe Rvine & -0.050 & 1.475 & -0.034 & 2.964 & -0.017 & 72.657 & 0.026 & 0.906 & 0.028 & 1.712 & 0.015 & 170.395 \\
				Gumbel & -0.048 & 1.556 & -0.031 & 3.096 & -0.016 & 73.027 & 0.024 & 0.973 & 0.024 & 1.849 & 0.013 & 159.284 \\
				Gumbel Cvine & -0.051 & 1.454 & -0.035 & 2.900 & -0.018 & 72.610 & 0.027 & 0.896 & 0.030 & 1.684 & 0.016 & 175.267 \\
				Gumbel Dvine & -0.050 & 1.456 & -0.034 & 2.916 & -0.017 & 73.011 & 0.027 & 0.905 & 0.030 & 1.702 & 0.016 & 176.638 \\
				Gumbel Rvine & -0.052 & 1.467 & -0.035 & 2.938 & -0.018 & 72.142 & 0.025 & 0.904 & 0.027 & 1.706 & 0.014 & 166.716 \\
				Mixed Cvine & -0.052 & 1.449 & -0.036 & 2.900 & -0.018 & 72.339 & 0.027 & 0.895 & 0.030 & 1.684 & 0.016 & 175.156 \\
				Mixed Dvine & -0.049 & 1.457 & -0.034 & 2.923 & -0.017 & 73.074 & 0.028 & 0.903 & 0.031 & 1.702 & 0.017 & 181.686 \\
				Mixed Rvine & -0.053 & 1.448 & -0.036 & 2.889 & -0.018 & 71.898 & 0.026 & 0.895 & 0.029 & 1.686 & 0.015 & 171.478 \\
				\bottomrule
			\end{tabular}%
			\begin{tablenotes}
				\footnotesize
				\item Note: This table provides in-sample performance of copula-based CVaR portfolios strategies. Portfolio strategies are constructed by utilizing realized assets' returns and optimal weights over the targeted period. Optimal weights are obtained from simulated returns by estimating the conditional mean and volatility from AR-GARCH model in Section \ref{MarginalModeling} and estimating the joint distribution (including dependency structure) and drawing 10000 simulations from different simple and vine copulas. In mixed vines, the copula selection is based on AIC including all the families in Table \ref{Cop.Fam}. In Panel A, the sample period is from January 1, 2008 to December 31, 2009. Panel B reports the results from June 3, 2003 to December 31, 2012. All the performance measures are in percentage. STARR (mean to CVaR) ratio and CVaR are estimated empirically at 10\% level. The terminal wealth, based on a buy-and-hold strategy with \$100 investment at the beginning of each period, is also reported.
			\end{tablenotes}
	\end{threeparttable}}
\end{table}

\begin{table}[H]
	\centering
	\resizebox{\textwidth}{!}{
		\begin{threeparttable}
			\caption{In-Sample Performance of GMV Portfolio Strategies}\label{InS.GMV}
			\begin{tabular}{lcccccccccccc}
				\toprule
				Sample Period & \multicolumn{6}{l}{\textit{Panel A: 2008-2009}} & \multicolumn{6}{l}{\textit{Panel B: 2003-2012}} \\
				\midrule
				\multicolumn{1}{r}{\multirow{2}[2]{*}{Forecasting Model}} & \multicolumn{1}{r}{\multirow{2}[2]{*}{Mean}} & \multicolumn{1}{p{4.055em}}{St.} & \multicolumn{1}{r}{\multirow{2}[2]{*}{SR}} & \multicolumn{1}{r}{\multirow{2}[2]{*}{CVaR}} & \multicolumn{1}{r}{\multirow{2}[2]{*}{STARR}} & \multicolumn{1}{p{4.055em}}{Terminal} & \multicolumn{1}{r}{\multirow{2}[2]{*}{Mean}} & \multicolumn{1}{p{4.055em}}{St.} & \multicolumn{1}{r}{\multirow{2}[2]{*}{SR}} & \multicolumn{1}{r}{\multirow{2}[2]{*}{CVaR}} & \multicolumn{1}{r}{\multirow{2}[2]{*}{STARR}} & \multicolumn{1}{p{4.055em}}{Terminal} \\
				&       & \multicolumn{1}{p{4.055em}}{Deviation} &       &       &       & \multicolumn{1}{p{4.055em}}{Wealth} &       & \multicolumn{1}{p{4.055em}}{Deviation} &       &       &       & \multicolumn{1}{p{4.055em}}{Wealth} \\
				\midrule
				EQW   & -0.050 & 1.569 & -0.032 & 3.115 & -0.016 & 72.143 & 0.027 & 1.000 & 0.027 & 1.901 & 0.014 & 172.071 \\
				Historical & -0.058 & 1.422 & -0.041 & 2.863 & -0.020 & 70.141 & 0.027 & 0.870 & 0.031 & 1.642 & 0.016 & 177.730 \\
				Student$-t$ & -0.051 & 1.472 & -0.034 & 2.964 & -0.017 & 72.577 & 0.024 & 0.898 & 0.027 & 1.699 & 0.014 & 165.785 \\
				Student$-t$ Cvine & -0.051 & 1.468 & -0.035 & 2.957 & -0.017 & 72.481 & 0.025 & 0.897 & 0.028 & 1.697 & 0.015 & 168.126 \\
				Student$-t$ Dvine & -0.051 & 1.468 & -0.035 & 2.956 & -0.017 & 72.496 & 0.025 & 0.897 & 0.028 & 1.696 & 0.015 & 168.124 \\
				Student$-t$ Rvine & -0.051 & 1.467 & -0.035 & 2.955 & -0.017 & 72.297 & 0.025 & 0.897 & 0.028 & 1.697 & 0.015 & 168.505 \\
				Clayton & -0.050 & 1.552 & -0.032 & 3.098 & -0.016 & 72.190 & 0.023 & 0.971 & 0.023 & 1.849 & 0.012 & 155.594 \\
				Clayton Cvine & -0.052 & 1.475 & -0.035 & 2.967 & -0.017 & 72.172 & 0.024 & 0.902 & 0.027 & 1.707 & 0.014 & 164.462 \\
				Clayton Dvine & -0.051 & 1.477 & -0.035 & 2.971 & -0.017 & 72.350 & 0.025 & 0.908 & 0.027 & 1.719 & 0.014 & 165.672 \\
				Clayton Rvine & -0.053 & 1.469 & -0.036 & 2.956 & -0.018 & 71.709 & 0.025 & 0.900 & 0.028 & 1.702 & 0.015 & 169.047 \\
				Frank & -0.050 & 1.562 & -0.032 & 3.123 & -0.016 & 72.208 & 0.020 & 0.968 & 0.021 & 1.842 & 0.011 & 147.752 \\
				Frank Cvine & -0.051 & 1.470 & -0.034 & 2.962 & -0.017 & 72.556 & 0.025 & 0.899 & 0.028 & 1.701 & 0.015 & 168.751 \\
				Frank Dvine & -0.051 & 1.470 & -0.034 & 2.961 & -0.017 & 72.578 & 0.025 & 0.900 & 0.028 & 1.703 & 0.015 & 168.039 \\
				Frank Rvine & -0.051 & 1.471 & -0.035 & 2.963 & -0.017 & 72.372 & 0.025 & 0.900 & 0.028 & 1.703 & 0.015 & 167.956 \\
				Joe   & -0.050 & 1.553 & -0.033 & 3.103 & -0.016 & 72.122 & 0.022 & 0.967 & 0.022 & 1.842 & 0.012 & 151.809 \\
				Joe Cvine & -0.052 & 1.476 & -0.035 & 2.967 & -0.017 & 72.121 & 0.024 & 0.903 & 0.027 & 1.709 & 0.014 & 164.013 \\
				Joe Dvine & -0.051 & 1.479 & -0.034 & 2.973 & -0.017 & 72.356 & 0.024 & 0.910 & 0.027 & 1.725 & 0.014 & 165.451 \\
				Joe Rvine & -0.052 & 1.479 & -0.035 & 2.979 & -0.017 & 72.148 & 0.025 & 0.907 & 0.027 & 1.718 & 0.014 & 167.263 \\
				Gumbel & -0.048 & 1.567 & -0.031 & 3.127 & -0.015 & 72.891 & 0.020 & 0.973 & 0.021 & 1.849 & 0.011 & 146.455 \\
				Gumbel Cvine & -0.051 & 1.467 & -0.035 & 2.952 & -0.017 & 72.306 & 0.025 & 0.897 & 0.027 & 1.695 & 0.015 & 166.527 \\
				Gumbel Dvine & -0.051 & 1.469 & -0.035 & 2.956 & -0.017 & 72.477 & 0.025 & 0.900 & 0.028 & 1.701 & 0.015 & 167.919 \\
				Gumbel Rvine & -0.052 & 1.468 & -0.035 & 2.957 & -0.017 & 72.172 & 0.025 & 0.898 & 0.028 & 1.699 & 0.015 & 167.506 \\
				Mixed Cvine & -0.051 & 1.466 & -0.035 & 2.952 & -0.017 & 72.489 & 0.025 & 0.896 & 0.028 & 1.693 & 0.015 & 168.675 \\
				Mixed Dvine & -0.051 & 1.466 & -0.035 & 2.951 & -0.017 & 72.461 & 0.025 & 0.897 & 0.028 & 1.695 & 0.015 & 168.404 \\
				Mixed Rvine & -0.051 & 1.465 & -0.035 & 2.950 & -0.017 & 72.279 & 0.025 & 0.895 & 0.028 & 1.692 & 0.015 & 168.710 \\
				\bottomrule
			\end{tabular}%
			\begin{tablenotes}
				\footnotesize
				\item Note: This table provides in-sample performance of copula-based GMV portfolios strategies. Portfolio strategies are constructed by utilizing realized assets' returns and optimal weights over the targeted period. Optimal weights are obtained from simulated returns by estimating the conditional mean and volatility from AR-GARCH model in Section \ref{MarginalModeling} and estimating the joint distribution (including dependency structure) and drawing 10000 simulations from different simple and vine copulas. In mixed vines, the copula selection is based on AIC including all the families in Table \ref{Cop.Fam}. In Panel A, the sample period is from January 1, 2008 to December 31, 2009. Panel B reports the results from June 3, 2003 to December 31, 2012. All the performance measures are in percentage. STARR (mean to CVaR) ratio and CVaR are estimated empirically at 10\% level. The terminal wealth, based on a buy-and-hold strategy with \$100 investment at the beginning of each period, is also reported.
			\end{tablenotes}
	\end{threeparttable}}
\end{table}

\begin{landscape}
\begin{table}[H]
	\centering
	\resizebox{1.4\textheight}{!}{
		\begin{threeparttable}
			\caption{Out-of-Sample Performance of SR Portfolio Strategies}\label{OutS.SR}
			\begin{tabular}{lllllllllllllllll}
				\toprule
				Sample Period & \multicolumn{8}{l}{\textit{Panel A: 2008-2009}}               & \multicolumn{8}{l}{\textit{Panel B: 2005-2012}} \\
				\midrule
				\multicolumn{1}{r}{\multirow{2}[2]{*}{Forecasting Model}} & \multicolumn{1}{r}{\multirow{2}[2]{*}{Mean}} & \multicolumn{1}{p{4.39em}}{St.} & \multicolumn{1}{r}{\multirow{2}[2]{*}{SR}} & \multicolumn{1}{r}{\multirow{2}[2]{*}{CVaR}} & \multicolumn{1}{r}{\multirow{2}[2]{*}{STARR}} & \multicolumn{1}{p{4.055em}}{Terminal} & \multicolumn{1}{p{6.89em}}{Terminal} & \multicolumn{1}{p{4.11em}}{Ave.} & \multicolumn{1}{r}{\multirow{2}[2]{*}{Mean}} & \multicolumn{1}{p{4.39em}}{St.} & \multicolumn{1}{r}{\multirow{2}[2]{*}{SR}} & \multicolumn{1}{r}{\multirow{2}[2]{*}{CVaR}} & \multicolumn{1}{r}{\multirow{2}[2]{*}{STARR}} & \multicolumn{1}{p{4.055em}}{Terminal} & \multicolumn{1}{p{6.89em}}{Terminal} & \multicolumn{1}{p{4.11em}}{Ave.} \\
				&       & \multicolumn{1}{p{4.39em}}{Deviation} &       &       &       & \multicolumn{1}{p{4.055em}}{Wealth} & \multicolumn{1}{p{6.89em}}{Wealth (TC=10)} & \multicolumn{1}{p{4.11em}}{Turnover} &       & \multicolumn{1}{p{4.39em}}{Deviation} &       &       &       & \multicolumn{1}{p{4.055em}}{Wealth} & \multicolumn{1}{p{6.89em}}{Wealth (TC=10)} & \multicolumn{1}{p{4.11em}}{Turnover} \\
				\midrule
				EQW   & -0.050 & 1.569 & -0.032 & 3.115 & -0.016 & 72.143 & 72.068 & 0.000 & 0.013 & 1.074 & 0.012 & 2.067 & 0.006 & 116.115 & 116.293 & 0.000 \\
				Historical & -0.188 & 2.082 & -0.090 & 4.416 & -0.043 & 33.284 & 29.525 & 0.228 & -0.035 & 1.366 & -0.026 & 2.688 & -0.013 & 40.187 & 31.288 & 0.132 \\
				Student$-t$ & 0.280 & 1.855 & 0.151 & 2.976 & 0.094 & 393.879 & 207.074 & 1.234 & 0.149 & 1.202 & 0.124 & 1.966 & 0.076 & 1700.662 & 176.599 & 1.131 \\
				Student$-t$ Cvine & 0.296 & 1.773 & 0.167 & 2.822 & 0.105 & 431.574 & 225.399 & 1.246 & 0.149 & 1.173 & 0.127 & 1.915 & 0.078 & 1716.394 & 186.304 & 1.109 \\
				Student$-t$ Dvine & 0.297 & 1.761 & 0.169 & 2.807 & 0.106 & 433.456 & 226.491 & 1.245 & 0.149 & 1.168 & 0.128 & 1.907 & 0.078 & 1729.172 & 189.888 & 1.103 \\
				Student$-t$ Rvine & 0.256 & 1.767 & 0.145 & 2.873 & 0.089 & 349.265 & 186.569 & 1.203 & 0.140 & 1.161 & 0.121 & 1.938 & 0.072 & 1435.665 & 161.186 & 1.092 \\
				Clayton & 0.305 & 1.783 & 0.171 & 2.762 & 0.111 & 452.633 & 241.262 & 1.204 & 0.154 & 1.179 & 0.131 & 1.919 & 0.080 & 1901.701 & 200.229 & 1.124 \\
				Clayton Cvine & 0.291 & 1.734 & 0.168 & 2.755 & 0.106 & 422.579 & 220.110 & 1.250 & 0.143 & 1.148 & 0.125 & 1.882 & 0.076 & 1537.723 & 178.190 & 1.076 \\
				Clayton Dvine & 0.281 & 1.726 & 0.163 & 2.790 & 0.101 & 401.218 & 210.121 & 1.240 & 0.139 & 1.145 & 0.122 & 1.888 & 0.074 & 1422.067 & 166.297 & 1.072 \\
				Clayton Rvine & 0.251 & 1.755 & 0.143 & 2.896 & 0.087 & 341.874 & 183.659 & 1.191 & 0.130 & 1.149 & 0.113 & 1.939 & 0.067 & 1175.745 & 140.814 & 1.060 \\
				Frank & 0.306 & 1.787 & 0.171 & 2.838 & 0.108 & 453.558 & 240.747 & 1.212 & 0.160 & 1.183 & 0.136 & 1.930 & 0.083 & 2139.180 & 216.130 & 1.145 \\
				Frank Cvine & 0.293 & 1.748 & 0.167 & 2.792 & 0.105 & 424.636 & 222.105 & 1.242 & 0.146 & 1.158 & 0.126 & 1.896 & 0.077 & 1608.106 & 184.041 & 1.082 \\
				Frank Dvine & 0.288 & 1.740 & 0.166 & 2.799 & 0.103 & 415.395 & 217.682 & 1.239 & 0.143 & 1.154 & 0.124 & 1.897 & 0.075 & 1520.465 & 177.362 & 1.073 \\
				Frank Rvine & 0.262 & 1.774 & 0.148 & 2.903 & 0.090 & 361.240 & 193.908 & 1.192 & 0.140 & 1.160 & 0.120 & 1.932 & 0.072 & 1427.340 & 169.665 & 1.063 \\
				Joe   & 0.312 & 1.787 & 0.174 & 2.762 & 0.113 & 468.037 & 249.200 & 1.207 & 0.160 & 1.200 & 0.134 & 1.964 & 0.082 & 2135.038 & 200.003 & 1.182 \\
				Joe Cvine & 0.282 & 1.716 & 0.164 & 2.742 & 0.103 & 403.050 & 211.184 & 1.238 & 0.139 & 1.137 & 0.122 & 1.877 & 0.074 & 1402.829 & 167.634 & 1.061 \\
				Joe Dvine & 0.277 & 1.715 & 0.162 & 2.781 & 0.100 & 393.005 & 207.216 & 1.227 & 0.137 & 1.138 & 0.120 & 1.883 & 0.073 & 1362.238 & 160.319 & 1.068 \\
				Joe Rvine & 0.258 & 1.745 & 0.148 & 2.862 & 0.090 & 355.676 & 191.142 & 1.190 & 0.133 & 1.141 & 0.116 & 1.924 & 0.069 & 1245.626 & 150.025 & 1.057 \\
				Gumbel & 0.336 & 1.801 & 0.187 & 2.789 & 0.121 & 531.961 & 276.554 & 1.252 & 0.168 & 1.196 & 0.140 & 1.911 & 0.088 & 2475.831 & 233.893 & 1.178 \\
				Gumbel Cvine & 0.289 & 1.742 & 0.166 & 2.780 & 0.104 & 416.397 & 217.151 & 1.249 & 0.147 & 1.157 & 0.127 & 1.896 & 0.077 & 1638.232 & 181.154 & 1.100 \\
				Gumbel Dvine & 0.291 & 1.778 & 0.164 & 2.835 & 0.103 & 420.168 & 219.495 & 1.245 & 0.148 & 1.171 & 0.126 & 1.909 & 0.077 & 1673.576 & 184.339 & 1.101 \\
				Gumbel Rvine & 0.272 & 1.731 & 0.157 & 2.765 & 0.098 & 381.866 & 203.106 & 1.211 & 0.140 & 1.141 & 0.123 & 1.904 & 0.074 & 1456.101 & 166.241 & 1.083 \\
				Mixed Cvine & 0.288 & 1.778 & 0.162 & 2.840 & 0.101 & 413.726 & 216.562 & 1.242 & 0.145 & 1.174 & 0.124 & 1.924 & 0.076 & 1595.079 & 173.658 & 1.107 \\
				Mixed Dvine & 0.299 & 1.781 & 0.168 & 2.821 & 0.106 & 437.107 & 227.633 & 1.252 & 0.150 & 1.177 & 0.128 & 1.913 & 0.078 & 1752.662 & 190.283 & 1.109 \\
				Mixed Rvine & 0.256 & 1.744 & 0.147 & 2.822 & 0.091 & 350.004 & 186.437 & 1.209 & 0.138 & 1.148 & 0.120 & 1.917 & 0.072 & 1371.709 & 153.655 & 1.093 \\
				\bottomrule
			\end{tabular}%
			\begin{tablenotes}
				\footnotesize
				\item Notes: This table presents the out-of-sample performance of SR portfolio strategies from using several vine and simple copulas. The two benchmarks are EQW portfolio and historical SR strategy. The performance measures include average return, volatility, the SR, CVaR, mean to CVaR ratio (STARR), and the final value of portfolio accumulation wealth (terminal wealth) assuming \$100 initial investment and average level of asset trade (average turnover). Terminal wealth is calculated for different levels (0 and 10 basis points) of the proportional TCs (see \cite{demiguel2009generalized}). The results are obtained using rolling window estimation, where the window is fixed and includes the last 500 daily returns for each stock market. In Panel A, the out-of-sample period is from January 1, 2008 to December 31, 2009. Panel B reports the result from May 3, 2005, to December 31, 2012.
			\end{tablenotes}
	\end{threeparttable}}
\end{table}
\end{landscape}
\begin{landscape}
\begin{table}[H]
	\centering
	\resizebox{1.4\textheight}{!}{
		\begin{threeparttable}
			\caption{Out-of-Sample Performance of CVaR Portfolio Strategies}\label{OutS.CVaR}
			\begin{tabular}{lcccccccccccccccc}
				\toprule
				Sample Period & \multicolumn{8}{l}{\textit{Panel A: 2008-2009}}               & \multicolumn{8}{l}{\textit{Panel B: 2005-2012}} \\
				\midrule
				\multicolumn{1}{r}{\multirow{2}[2]{*}{Forecasting Model}} & \multicolumn{1}{r}{\multirow{2}[2]{*}{Mean}} & \multicolumn{1}{p{4.39em}}{St.} & \multicolumn{1}{r}{\multirow{2}[2]{*}{SR}} & \multicolumn{1}{r}{\multirow{2}[2]{*}{CVaR}} & \multicolumn{1}{r}{\multirow{2}[2]{*}{STARR}} & \multicolumn{1}{p{4.055em}}{Terminal} & \multicolumn{1}{p{6.89em}}{Terminal} & \multicolumn{1}{p{4.11em}}{Ave.} & \multicolumn{1}{r}{\multirow{2}[2]{*}{Mean}} & \multicolumn{1}{p{4.39em}}{St.} & \multicolumn{1}{r}{\multirow{2}[2]{*}{SR}} & \multicolumn{1}{r}{\multirow{2}[2]{*}{CVaR}} & \multicolumn{1}{r}{\multirow{2}[2]{*}{STARR}} & \multicolumn{1}{p{4.055em}}{Terminal} & \multicolumn{1}{p{6.89em}}{Terminal} & \multicolumn{1}{p{4.11em}}{Ave.} \\
				&       & \multicolumn{1}{p{4.39em}}{Deviation} &       &       &       & \multicolumn{1}{p{4.055em}}{Wealth} & \multicolumn{1}{p{6.89em}}{Wealth (TC=10)} & \multicolumn{1}{p{4.11em}}{Turnover} &       & \multicolumn{1}{p{4.39em}}{Deviation} &       &       &       & \multicolumn{1}{p{4.055em}}{Wealth} & \multicolumn{1}{p{6.89em}}{Wealth (TC=10)} & \multicolumn{1}{p{4.11em}}{Turnover} \\
				\midrule
				EQW   & -0.050 & 1.569 & -0.032 & 3.115 & -0.016 & 72.143 & 72.068 & 0.000 & 0.013 & 1.074 & 0.012 & 2.067 & 0.006 & 116.115 & 116.293 & 0.000 \\
				Historical & -0.109 & 1.593 & -0.068 & 3.230 & -0.034 & 52.927 & 52.376 & 0.019 & -0.012 & 1.034 & -0.011 & 1.988 & -0.006 & 70.664 & 68.694 & 0.016 \\
				Student$-t$ & -0.105 & 1.607 & -0.066 & 3.150 & -0.033 & 53.908 & 43.549 & 0.400 & -0.021 & 1.023 & -0.021 & 1.926 & -0.011 & 58.873 & 26.762 & 0.394 \\
				Student$-t$ Cvine & -0.097 & 1.546 & -0.063 & 3.070 & -0.032 & 56.456 & 48.503 & 0.285 & -0.015 & 0.990 & -0.016 & 1.860 & -0.008 & 66.265 & 38.488 & 0.272 \\
				Student$-t$ Dvine & -0.092 & 1.550 & -0.060 & 3.051 & -0.030 & 57.915 & 49.743 & 0.285 & -0.014 & 0.992 & -0.014 & 1.859 & -0.008 & 67.853 & 39.290 & 0.273 \\
				Student$-t$ Rvine & -0.106 & 1.580 & -0.067 & 3.137 & -0.034 & 53.906 & 46.252 & 0.283 & -0.020 & 1.005 & -0.020 & 1.899 & -0.011 & 60.272 & 33.157 & 0.299 \\
				Clayton & -0.089 & 1.732 & -0.051 & 3.352 & -0.027 & 58.165 & 46.625 & 0.414 & -0.021 & 1.100 & -0.019 & 2.061 & -0.010 & 58.254 & 27.663 & 0.372 \\
				Clayton Cvine & -0.103 & 1.599 & -0.065 & 3.145 & -0.033 & 54.535 & 46.338 & 0.302 & -0.016 & 1.016 & -0.015 & 1.907 & -0.008 & 65.587 & 36.148 & 0.298 \\
				Clayton Dvine & -0.093 & 1.565 & -0.059 & 3.073 & -0.030 & 57.803 & 48.798 & 0.318 & -0.013 & 1.003 & -0.013 & 1.889 & -0.007 & 69.827 & 37.775 & 0.308 \\
				Clayton Rvine & -0.085 & 1.606 & -0.053 & 3.149 & -0.027 & 59.942 & 49.256 & 0.368 & -0.010 & 1.020 & -0.010 & 1.918 & -0.005 & 73.372 & 36.729 & 0.346 \\
				Frank & -0.063 & 1.619 & -0.039 & 3.170 & -0.020 & 67.161 & 61.076 & 0.178 & -0.002 & 1.061 & -0.002 & 2.001 & -0.001 & 85.917 & 60.979 & 0.172 \\
				Frank Cvine & -0.074 & 1.541 & -0.048 & 3.000 & -0.025 & 63.908 & 55.995 & 0.247 & -0.007 & 0.994 & -0.007 & 1.856 & -0.004 & 78.226 & 48.374 & 0.241 \\
				Frank Dvine & -0.076 & 1.542 & -0.049 & 3.019 & -0.025 & 63.140 & 55.147 & 0.253 & -0.007 & 0.995 & -0.007 & 1.865 & -0.004 & 78.129 & 48.483 & 0.239 \\
				Frank Rvine & -0.076 & 1.557 & -0.049 & 3.040 & -0.025 & 63.252 & 55.501 & 0.244 & -0.008 & 0.999 & -0.008 & 1.872 & -0.004 & 77.600 & 46.728 & 0.254 \\
				Joe   & -0.058 & 1.578 & -0.037 & 3.127 & -0.019 & 69.132 & 64.275 & 0.136 & -0.001 & 1.048 & -0.001 & 1.993 & 0.000 & 88.457 & 65.979 & 0.147 \\
				Joe Cvine & -0.099 & 1.603 & -0.062 & 3.142 & -0.032 & 55.689 & 47.164 & 0.308 & -0.015 & 1.017 & -0.014 & 1.906 & -0.008 & 66.981 & 36.400 & 0.305 \\
				Joe Dvine & -0.093 & 1.569 & -0.059 & 3.064 & -0.030 & 57.594 & 48.586 & 0.320 & -0.012 & 1.009 & -0.012 & 1.891 & -0.007 & 70.339 & 36.995 & 0.321 \\
				Joe Rvine & -0.102 & 1.602 & -0.064 & 3.155 & -0.032 & 54.953 & 44.862 & 0.380 & -0.016 & 1.022 & -0.016 & 1.937 & -0.008 & 64.573 & 31.736 & 0.356 \\
				Gumbel & -0.066 & 1.623 & -0.040 & 3.160 & -0.021 & 66.206 & 59.397 & 0.204 & 0.000 & 1.045 & 0.000 & 1.962 & 0.000 & 88.937 & 60.654 & 0.192 \\
				Gumbel Cvine & -0.092 & 1.595 & -0.057 & 3.079 & -0.030 & 58.095 & 48.909 & 0.319 & -0.013 & 1.013 & -0.012 & 1.891 & -0.007 & 69.938 & 38.357 & 0.301 \\
				Gumbel Dvine & -0.087 & 1.592 & -0.055 & 3.074 & -0.028 & 59.406 & 49.988 & 0.320 & -0.012 & 1.010 & -0.012 & 1.876 & -0.006 & 71.215 & 38.143 & 0.312 \\
				Gumbel Rvine & -0.096 & 1.598 & -0.060 & 3.132 & -0.031 & 56.725 & 47.288 & 0.338 & -0.014 & 1.014 & -0.014 & 1.894 & -0.008 & 67.484 & 34.112 & 0.342 \\
				Mixed Cvine & -0.101 & 1.564 & -0.064 & 3.089 & -0.033 & 55.508 & 47.497 & 0.289 & -0.015 & 1.000 & -0.015 & 1.875 & -0.008 & 66.618 & 37.606 & 0.286 \\
				Mixed Dvine & -0.090 & 1.594 & -0.056 & 3.103 & -0.029 & 58.660 & 49.554 & 0.312 & -0.014 & 1.011 & -0.013 & 1.875 & -0.007 & 68.621 & 37.932 & 0.297 \\
				Mixed Rvine & -0.092 & 1.601 & -0.057 & 3.113 & -0.030 & 57.969 & 49.075 & 0.308 & -0.017 & 1.014 & -0.017 & 1.887 & -0.009 & 64.202 & 33.840 & 0.321 \\
				\bottomrule
			\end{tabular}%
			\begin{tablenotes}
				\footnotesize
				\item Notes: This table presents the out-of-sample performance of CVaR portfolio strategies from using several vine and simple copulas. The two benchmarks are EQW portfolio and historical CVaR strategy. The performance measures include average return, volatility, the SR, CVaR, mean to CVaR ratio (STARR), and the final value of portfolio accumulation wealth (terminal wealth) assuming \$100 initial investment and average level of asset trade (average turnover). Terminal wealth is calculated for different levels (0 and 10 basis points) of the proportional TCs (see \cite{demiguel2009generalized}). The results are obtained using rolling window estimation, where the window is fixed and includes the last 500 daily returns for each stock market. In Panel A, the out-of-sample period is from January 1, 2008 to December 31, 2009. Panel B reports the result from May 3, 2005, to December 31, 2012.
			\end{tablenotes}
	\end{threeparttable}}
\end{table}
\end{landscape}
\begin{landscape}
\begin{table}[H]
	\centering
	\resizebox{1.4\textheight}{!}{
		\begin{threeparttable}
			\caption{Out-of-Sample Performance of MV Portfolio Strategies}\label{OutS.MV}
			\begin{tabular}{lcccccccccccccccc}
				\toprule
				Sample Period & \multicolumn{8}{l}{\textit{Panel A: 2008-2009}}               & \multicolumn{8}{l}{\textit{Panel B: 2005-2012}} \\
				\midrule
				\multicolumn{1}{r}{\multirow{2}[2]{*}{Forecasting Model}} & \multicolumn{1}{r}{\multirow{2}[2]{*}{Mean}} & \multicolumn{1}{p{4.39em}}{St.} & \multicolumn{1}{r}{\multirow{2}[2]{*}{SR}} & \multicolumn{1}{r}{\multirow{2}[2]{*}{CVaR}} & \multicolumn{1}{r}{\multirow{2}[2]{*}{STARR}} & \multicolumn{1}{p{4.055em}}{Terminal} & \multicolumn{1}{p{6.89em}}{Terminal} & \multicolumn{1}{p{4.11em}}{Ave.} & \multicolumn{1}{r}{\multirow{2}[2]{*}{Mean}} & \multicolumn{1}{p{4.39em}}{St.} & \multicolumn{1}{r}{\multirow{2}[2]{*}{SR}} & \multicolumn{1}{r}{\multirow{2}[2]{*}{CVaR}} & \multicolumn{1}{r}{\multirow{2}[2]{*}{STARR}} & \multicolumn{1}{p{4.055em}}{Terminal} & \multicolumn{1}{p{6.89em}}{Terminal} & \multicolumn{1}{p{4.11em}}{Ave.} \\
				&       & \multicolumn{1}{p{4.39em}}{Deviation} &       &       &       & \multicolumn{1}{p{4.055em}}{Wealth} & \multicolumn{1}{p{6.89em}}{Wealth (TC=10)} & \multicolumn{1}{p{4.11em}}{Turnover} &       & \multicolumn{1}{p{4.39em}}{Deviation} &       &       &       & \multicolumn{1}{p{4.055em}}{Wealth} & \multicolumn{1}{p{6.89em}}{Wealth (TC=10)} & \multicolumn{1}{p{4.11em}}{Turnover} \\
				\midrule
				EQW   & -0.050 & 1.569 & -0.032 & 3.115 & -0.016 & 72.143 & 72.133 & 0.000 & 0.013 & 1.074 & 0.012 & 2.067 & 0.006 & 116.115 & 116.293 & 0.000 \\
				Historical & -0.074 & 1.466 & -0.050 & 2.952 & -0.025 & 64.168 & 64.088 & 0.018 & -0.002 & 0.952 & -0.002 & 1.832 & -0.001 & 88.101 & 85.981 & 0.014 \\
				Student$-t$ & -0.117 & 1.597 & -0.073 & 3.164 & -0.037 & 50.894 & 49.964 & 0.270 & -0.022 & 1.020 & -0.022 & 1.926 & -0.012 & 57.149 & 33.287 & 0.272 \\
				Student$-t$ Cvine & -0.115 & 1.591 & -0.072 & 3.154 & -0.036 & 51.409 & 50.511 & 0.257 & -0.022 & 1.016 & -0.021 & 1.916 & -0.011 & 58.119 & 34.778 & 0.258 \\
				Student$-t$ Dvine & -0.115 & 1.591 & -0.072 & 3.148 & -0.036 & 51.490 & 50.593 & 0.257 & -0.022 & 1.016 & -0.021 & 1.916 & -0.011 & 58.179 & 34.816 & 0.258 \\
				Student$-t$ Rvine & -0.114 & 1.593 & -0.071 & 3.151 & -0.036 & 51.794 & 50.867 & 0.257 & -0.022 & 1.017 & -0.021 & 1.919 & -0.011 & 58.291 & 34.693 & 0.261 \\
				Clayton & -0.104 & 1.698 & -0.062 & 3.313 & -0.032 & 53.750 & 52.762 & 0.290 & -0.023 & 1.085 & -0.021 & 2.052 & -0.011 & 55.933 & 33.710 & 0.253 \\
				Clayton Cvine & -0.108 & 1.591 & -0.068 & 3.134 & -0.035 & 53.152 & 52.238 & 0.260 & -0.019 & 1.014 & -0.019 & 1.914 & -0.010 & 61.396 & 37.498 & 0.247 \\
				Clayton Dvine & -0.111 & 1.580 & -0.070 & 3.117 & -0.036 & 52.440 & 51.627 & 0.248 & -0.019 & 1.011 & -0.019 & 1.907 & -0.010 & 61.118 & 37.765 & 0.242 \\
				Clayton Rvine & -0.107 & 1.590 & -0.067 & 3.118 & -0.034 & 53.605 & 52.671 & 0.257 & -0.018 & 1.014 & -0.018 & 1.907 & -0.009 & 62.869 & 38.489 & 0.246 \\
				Frank & -0.108 & 1.697 & -0.064 & 3.321 & -0.033 & 52.788 & 51.818 & 0.287 & -0.025 & 1.088 & -0.023 & 2.059 & -0.012 & 54.024 & 31.591 & 0.269 \\
				Frank Cvine & -0.113 & 1.590 & -0.071 & 3.146 & -0.036 & 51.860 & 51.002 & 0.248 & -0.021 & 1.016 & -0.021 & 1.915 & -0.011 & 58.722 & 35.809 & 0.248 \\
				Frank Dvine & -0.112 & 1.590 & -0.071 & 3.139 & -0.036 & 52.106 & 51.261 & 0.246 & -0.021 & 1.016 & -0.020 & 1.916 & -0.011 & 59.231 & 36.413 & 0.244 \\
				Frank Rvine & -0.112 & 1.592 & -0.070 & 3.138 & -0.036 & 52.245 & 51.379 & 0.246 & -0.020 & 1.016 & -0.020 & 1.914 & -0.011 & 59.782 & 36.377 & 0.250 \\
				Joe   & -0.106 & 1.689 & -0.063 & 3.307 & -0.032 & 53.449 & 52.517 & 0.271 & -0.025 & 1.095 & -0.023 & 2.074 & -0.012 & 53.616 & 30.620 & 0.281 \\
				Joe Cvine & -0.109 & 1.588 & -0.069 & 3.120 & -0.035 & 52.921 & 52.057 & 0.252 & -0.019 & 1.012 & -0.019 & 1.909 & -0.010 & 61.241 & 37.864 & 0.241 \\
				Joe Dvine & -0.109 & 1.572 & -0.069 & 3.091 & -0.035 & 53.001 & 52.202 & 0.240 & -0.019 & 1.009 & -0.019 & 1.899 & -0.010 & 61.633 & 38.195 & 0.240 \\
				Joe Rvine & -0.109 & 1.585 & -0.069 & 3.121 & -0.035 & 53.125 & 52.221 & 0.251 & -0.018 & 1.011 & -0.018 & 1.902 & -0.010 & 62.163 & 38.391 & 0.242 \\
				Gumbel & -0.110 & 1.695 & -0.065 & 3.330 & -0.033 & 52.278 & 51.305 & 0.289 & -0.021 & 1.072 & -0.019 & 2.028 & -0.010 & 58.684 & 34.281 & 0.269 \\
				Gumbel Cvine & -0.111 & 1.592 & -0.070 & 3.142 & -0.035 & 52.336 & 51.420 & 0.260 & -0.020 & 1.016 & -0.020 & 1.911 & -0.011 & 59.677 & 36.023 & 0.254 \\
				Gumbel Dvine & -0.114 & 1.595 & -0.072 & 3.145 & -0.036 & 51.538 & 50.659 & 0.260 & -0.021 & 1.016 & -0.021 & 1.912 & -0.011 & 58.450 & 35.196 & 0.255 \\
				Gumbel Rvine & -0.115 & 1.596 & -0.072 & 3.155 & -0.037 & 51.299 & 50.383 & 0.258 & -0.021 & 1.016 & -0.020 & 1.914 & -0.011 & 59.147 & 35.523 & 0.256 \\
				Mixed Cvine & -0.112 & 1.584 & -0.071 & 3.126 & -0.036 & 52.131 & 51.211 & 0.262 & -0.021 & 1.014 & -0.020 & 1.907 & -0.011 & 59.413 & 35.544 & 0.258 \\
				Mixed Dvine & -0.115 & 1.596 & -0.072 & 3.160 & -0.037 & 51.235 & 50.341 & 0.260 & -0.022 & 1.020 & -0.021 & 1.921 & -0.011 & 57.980 & 34.532 & 0.260 \\
				Mixed Rvine & -0.111 & 1.590 & -0.070 & 3.139 & -0.036 & 52.360 & 51.450 & 0.255 & -0.020 & 1.015 & -0.020 & 1.908 & -0.011 & 59.776 & 35.520 & 0.262 \\
				\bottomrule
			\end{tabular}%
			\begin{tablenotes}
				\footnotesize
				\item Notes: This table presents the out-of-sample performance of MV portfolio strategies from using several vine and simple copulas. The two benchmarks are EQW portfolio and historical MV strategy. The performance measures include average return, volatility, the SR, CVaR, mean to CVaR ratio (STARR), and the final value of portfolio accumulation wealth (terminal wealth) assuming \$100 initial investment and average level of asset trade (average turnover). Terminal wealth is calculated for different levels (0 and 10 basis points) of the proportional TCs (see \cite{demiguel2009generalized}). The results are obtained using rolling window estimation, where the window is fixed and includes the last 500 daily returns for each stock market. In Panel A, the out-of-sample period is from January 1, 2008 to December 31, 2009. Panel B reports the result from May 3, 2005, to December 31, 2012.
			\end{tablenotes}
	\end{threeparttable}}
\end{table}
\end{landscape}

\begin{landscape}
\begin{table}[H]
	\centering
	\resizebox{1.4\textheight}{!}{
		\begin{threeparttable}
			\caption{VaR Back-testing}\label{VaRB}
			\begin{tabular}{lccccccccccccccccccccc}
				\toprule
				\multicolumn{1}{l}{\multirow{2}[4]{*}{Forecasting Model}} & \multicolumn{7}{l}{SR Portfolio Strategies}           & \multicolumn{7}{l}{CVaR Portfolio Strategies}         & \multicolumn{7}{l}{GMV Portfolio Strategies} \\
				\cmidrule{2-22}          & NE    & UC    & CC    & DQ    & AD    & AE    & AQL   & NE    & UC    & CC    & DQ    & AD    & AE    & AQL   & NE    & UC    & CC    & DQ    & AD    & AE    & AQL \\
				\midrule
				\multicolumn{22}{p{99.23em}}{\textit{Panel A: 2007-2009}} \\
				Student$-t$ & 21    & 15.29*** & 15.58*** & 68.07*** & 0.897 & 2.679 & 0.057 & 19    & 11.48*** & 12.05*** & 46.62*** & 0.852 & 2.423 & 0.053 & 14    & 3.96** & \textbf{4.44} & 45.92*** & 0.791 & 1.786 & 0.046 \\
				Student$-t$ Cvine & 28    & 31.49*** & 34.39*** & 135.19*** & 0.829 & 3.571 & 0.06  & 16    & 6.59** & 7.64** & 55.63*** & 0.931 & 2.041 & 0.049 & 16    & 6.60** & 7.22** & 44.80*** & 0.832 & 2.041 & 0.047 \\
				Student$-t$ Dvine & 28    & 31.49*** & 34.39*** & 135.25*** & 0.839 & 3.571 & 0.061 & 14    & 3.96** & \textbf{4.44} & 23.74*** & 0.949 & 1.786 & 0.047 & 15    & 5.21** & \textbf{5.76*} & 26.35*** & 0.889 & 1.913 & 0.047 \\
				Student$-t$ Rvine & 29    & 34.13*** & 34.89*** & 136.18*** & 0.751 & 3.699 & 0.058 & 19    & 11.48*** & 12.04*** & 79.51*** & 0.752 & 2.423 & 0.048 & 17    & 8.10*** & 8.97** & 43.68*** & 0.783 & 2.168 & 0.047 \\
				Clayton & 16    & 6.59** & 7.54*** & 31.79*** & 0.900 & 2.041 & 0.053 & 18    & 9.73*** & 10.53*** & 70.34*** & 0.724 & 2.296 & 0.051 & 11    & \textbf{1.14} & \textbf{1.43} & 14.13** & 0.345 & 1.403 & 0.039 \\
				Clayton Cvine & 20    & 13.33*** & 16.05*** & 70.80*** & 0.805 & 2.551 & 0.052 & 25    & 24.04*** & 24.11*** & 132.38*** & 0.849 & 3.189 & 0.058 & 14    & 3.96** & 5.44** & 43.41*** & 0.635 & 1.786 & 0.043 \\
				Clayton Dvine & 23    & 19.49*** & 21.34*** & 92.76*** & 0.782 & 2.934 & 0.054 & 21    & 15.29*** & 17.85*** & 111.28*** & 0.863 & 2.679 & 0.053 & 13    & \textbf{2.86*} & \textbf{3.27} & 20.82*** & 0.724 & 1.658 & 0.042 \\
				Clayton Rvine & 21    & 15.29*** & 15.58*** & 58.68*** & 0.755 & 2.679 & 0.052 & 20    & 13.33*** & 13.78*** & 52.43*** & 0.823 & 2.551 & 0.053 & 13    & \textbf{2.86*} & \textbf{3.27} & 21.28*** & 0.717 & 1.658 & 0.044 \\
				Frank & 52    & 111.01*** & 111.72*** & 381.61*** & 0.837 & 6.633 & 0.076 & 14    & 3.96** & \textbf{4.47} & 40.39*** & 0.401 & 1.786 & 0.028 & 33    & 45.36*** & 45.47*** & 166.16*** & 0.547 & 4.209 & 0.044 \\
				Frank Cvine & 32    & 42.46*** & 44.25*** & 151.69*** & 0.882 & 4.082 & 0.062 & 14    & 3.96** & \textbf{4.44} & 38.76*** & 0.572 & 1.786 & 0.036 & 22    & 17.34*** & 17.59*** & 79.29*** & 0.642 & 2.806 & 0.044 \\
				Frank Dvine & 32    & 42.46*** & 44.25*** & 151.70*** & 0.882 & 4.082 & 0.062 & 13    & \textbf{2.86*} & \textbf{3.27} & 22.25*** & 0.610 & 1.658 & 0.036 & 27    & 28.93*** & 30.03*** & 108.75*** & 0.557 & 3.444 & 0.045 \\
				Frank Rvine & 37    & 57.61*** & 58.44*** & 194.94*** & 0.748 & 4.719 & 0.061 & 16    & 6.59** & 7.22** & 67.97*** & 0.618 & 2.041 & 0.038 & 26    & 26.45*** & 27.74*** & 115.45*** & 0.604 & 3.316 & 0.045 \\
				Joe   & 79    & 229.43*** & 230.73*** & 863.81*** & 0.851 & 10.077 & 0.102 & 21    & 15.29*** & 16.44*** & 81.31*** & 0.373 & 2.679 & 0.026 & 48    & 95.74*** & 95.75*** & 458.00*** & 0.569 & 6.122 & 0.051 \\
				Joe Cvine & 22    & 17.34*** & 19.46*** & 75.25*** & 0.707 & 2.806 & 0.051 & 25    & 24.04*** & 25.55*** & 108.63*** & 0.907 & 3.189 & 0.060 & 14    & 3.96** & \textbf{4.44} & 20.75*** & 0.632 & 1.786 & 0.042 \\
				Joe Dvine & 23    & 19.49*** & 21.34*** & 76.01*** & 0.790 & 2.934 & 0.054 & 23    & 19.49*** & 19.66*** & 145.35*** & 0.793 & 2.934 & 0.053 & 13    & \textbf{2.86*} & \textbf{3.27} & 20.59*** & 0.669 & 1.658 & 0.041 \\
				Joe Rvine & 20    & 13.33*** & 13.72*** & 56.51*** & 0.794 & 2.551 & 0.053 & 23    & 19.49*** & 21.47*** & 69.06*** & 0.772 & 2.934 & 0.055 & 12    & \textbf{1.92} & \textbf{2.26} & 20.95*** & 0.788 & 1.531 & 0.044 \\
				Gumbel & 47    & 92.04*** & 92.54*** & 310.16*** & 0.779 & 5.995 & 0.069 & 15    & 5.21** & \textbf{5.80*} & 38.94*** & 0.447 & 1.913 & 0.031 & 33    & 45.36*** & 45.47*** & 184.08*** & 0.531 & 4.209 & 0.045 \\
				Gumbel Cvine & 21    & 15.29*** & 17.69*** & 90.13*** & 0.904 & 2.679 & 0.057 & 18    & 9.73*** & 10.53*** & 61.96*** & 1.110 & 2.296 & 0.058 & 13    & \textbf{2.86*} & \textbf{3.27} & 21.48*** & 0.859 & 1.658 & 0.046 \\
				Gumbel Dvine & 24    & 21.72*** & 26.09*** & 110.52*** & 0.825 & 3.061 & 0.057 & 21    & 15.29*** & 15.63*** & 79.79*** & 0.867 & 2.679 & 0.055 & 13    & \textbf{2.86*} & \textbf{3.27} & 21.26*** & 0.801 & 1.658 & 0.045 \\
				Gumbel Rvine & 24    & 21.72*** & 21.81*** & 80.25*** & 0.791 & 3.061 & 0.056 & 21    & 15.29*** & 17.85*** & 63.92*** & 0.826 & 2.679 & 0.054 & 14    & 3.96** & \textbf{4.44} & 22.00*** & 0.796 & 1.786 & 0.046 \\
				Mixed Cvine & 25    & 24.04*** & 28.00*** & 115.57*** & 0.847 & 3.189 & 0.059 & 17    & 8.10*** & 8.81** & 49.39*** & 0.971 & 2.168 & 0.052 & 14    & 3.96** & \textbf{4.44} & 21.00*** & 0.926 & 1.786 & 0.047 \\
				Mixed Dvine & 23    & 19.49*** & 24.29*** & 107.84*** & 0.943 & 2.934 & 0.059 & 15    & 5.21** & \textbf{5.76*} & 27.22*** & 0.912 & 1.913 & 0.048 & 15    & 5.21** & \textbf{5.76*} & 25.04*** & 0.812 & 1.913 & 0.046 \\
				Mixed Rvine & 25    & 24.04*** & 25.44*** & 91.09*** & 0.816 & 3.189 & 0.057 & 15    & 5.21** & \textbf{5.76*} & 22.92*** & 0.783 & 1.913 & 0.046 & 15    & 5.21** & \textbf{5.76*} & 24.11*** & 0.812 & 1.913 & 0.046 \\
				\multicolumn{22}{l}{\textit{Panel B: 2005-2012}} \\
				Student$-t$ & 26    & \textbf{1.66} & \textbf{5.67*} & 34.36*** & 0.870 & 1.300 & 0.037 & 11    & 4.89** & 8.79** & 22.96*** & 2.159 & 0.550 & 0.036 & 15    & \textbf{1.38} & \textbf{4.08*} & \textbf{13.76*} & 1.512 & 0.750 & 0.036 \\
				Student$-t$ Cvine & 27    & \textbf{2.23*} & \textbf{5.97*} & 32.65*** & 0.821 & 1.350 & 0.036 & 12    & \textbf{3.77*} & 7.33** & 20.07*** & 1.856 & 0.600 & 0.034 & 19    & \textbf{0.05} & \textbf{1.91*} & 23.55*** & 1.290 & 0.950 & 0.035 \\
				Student$-t$ Dvine & 29    & \textbf{3.59*} & 6.85** & 32.06*** & 0.780 & 1.450 & 0.036 & 12    & \textbf{3.77*} & 7.33** & 20.11*** & 1.890 & 0.600 & 0.034 & 19    & \textbf{0.05} & \textbf{1.91*} & 23.54*** & 1.300 & 0.950 & 0.035 \\
				Student$-t$ Rvine & 30    & 4.38** & 7.41** & 63.31*** & 0.826 & 1.500 & 0.037 & 13    & \textbf{2.82*} & 6.07** & 40.87*** & 1.785 & 0.650 & 0.035 & 20    & \textbf{0.00} & \textbf{1.68} & 21.94*** & 1.223 & 1.000 & 0.035 \\
				Clayton & 26    & \textbf{1.66} & \textbf{2.54} & 49.34*** & 0.713 & 1.300 & 0.036 & 22    & \textbf{0.20} & \textbf{1.57} & 31.70*** & 1.294 & 1.100 & 0.039 & 23    & \textbf{0.43} & \textbf{1.67} & 30.20*** & 1.221 & 1.150 & 0.039 \\
				ClaytonCvine & 29    & \textbf{3.59*} & 6.85** & 47.96*** & 0.691 & 1.450 & 0.034 & 20    & \textbf{0.00} & \textbf{1.68} & 23.82*** & 1.243 & 1.000 & 0.035 & 21    & \textbf{0.05} & \textbf{5.62*} & 35.30*** & 1.206 & 1.050 & 0.035 \\
				Clayton Dvine & 36    & 10.45*** & 12.36*** & 43.75*** & 0.614 & 1.800 & 0.035 & 20    & \textbf{0.00} & \textbf{1.68} & 23.99*** & 1.215 & 1.000 & 0.035 & 23    & \textbf{0.43} & \textbf{5.33*} & 51.17*** & 1.090 & 1.150 & 0.035 \\
				Clayton Rvine & 30    & 4.38** & 7.41** & 48.39*** & 0.765 & 1.500 & 0.036 & 20    & \textbf{0.00} & 11.48*** & 47.87*** & 1.219 & 1.000 & 0.035 & 21    & \textbf{0.05} & \textbf{5.62*} & 35.34*** & 1.177 & 1.050 & 0.035 \\
				Frank & 103   & 175.16*** & 191.58*** & 541.06*** & 0.673 & 5.150 & 0.052 & 101   & 168.48*** & 174.34*** & 425.18*** & 0.662 & 5.050 & 0.049 & 113   & 209.79*** & 217.69*** & 564.23*** & 0.695 & 5.650 & 0.055 \\
				Frank Cvine & 49    & 30.24*** & 34.61*** & 156.27*** & 0.677 & 2.450 & 0.038 & 37    & 11.67*** & 11.8*** & 48.73*** & 0.809 & 1.850 & 0.035 & 40    & 15.65*** & 19.44*** & 73.61*** & 0.846 & 2.000 & 0.037 \\
				Frank Dvine & 49    & 30.24*** & 37.54*** & 167.41*** & 0.690 & 2.450 & 0.038 & 41    & 17.09*** & 23.76*** & 82.72*** & 0.737 & 2.050 & 0.035 & 43    & 20.1*** & 23.22*** & 83.56*** & 0.803 & 2.150 & 0.037 \\
				Frank Rvine & 51    & 33.97*** & 37.86*** & 107.55*** & 0.728 & 2.550 & 0.039 & 40    & 15.65*** & 19.44*** & 76.35*** & 0.780 & 2.000 & 0.035 & 44    & 21.68*** & 24.59*** & 73.01*** & 0.813 & 2.200 & 0.037 \\
				Joe   & 167   & 426.03*** & 436.34*** & 1265.24*** & 0.637 & 8.350 & 0.068 & 157   & 382.71*** & 403.13*** & 1194.61*** & 0.662 & 7.850 & 0.065 & 170   & 439.28*** & 451.75*** & 1406.75*** & 0.715 & 8.500 & 0.073 \\
				Joe Cvine & 31    & 5.23** & 16*** & 80.77*** & 0.664 & 1.550 & 0.034 & 22    & \textbf{0.20} & 10.52*** & 57.38*** & 1.164 & 1.100 & 0.035 & 20    & \textbf{0.00} & \textbf{1.68} & 23.20*** & 1.289 & 1.000 & 0.035 \\
				Joe Dvine & 34    & 8.18*** & 10.43*** & 45.25*** & 0.608 & 1.700 & 0.034 & 21    & \textbf{0.05} & \textbf{1.57} & 22.10*** & 1.179 & 1.050 & 0.035 & 24    & \textbf{0.76} & \textbf{5.34*} & 48.70*** & 1.049 & 1.200 & 0.035 \\
				Joe Rvine & 31    & 5.23** & 16*** & 82.65*** & 0.679 & 1.550 & 0.035 & 19    & \textbf{0.05} & 12.17*** & 51.32*** & 1.307 & 0.950 & 0.036 & 20    & \textbf{0.00} & \textbf{5.95*} & 37.65*** & 1.170 & 1.000 & 0.035 \\
				Gumbel & 88    & 127.12*** & 145.45*** & 430.63*** & 0.629 & 4.400 & 0.046 & 79    & 100.82*** & 104.78*** & 286.68*** & 0.657 & 3.950 & 0.043 & 85    & 118.13*** & 122.57*** & 334.05*** & 0.708 & 4.250 & 0.047 \\
				Gumbel Cvine & 25    & \textbf{1.17} & \textbf{5.46*} & 35.49*** & 0.722 & 1.250 & 0.035 & 13    & \textbf{2.82*} & 6.07** & 21.46*** & 1.775 & 0.650 & 0.036 & 16    & \textbf{0.87} & \textbf{3.33} & \textbf{12.23*} & 1.462 & 0.800 & 0.036 \\
				Gumbel Dvine & 30    & 4.38** & 7.41** & 29.59*** & 0.688 & 1.500 & 0.035 & 12    & \textbf{3.77*} & 7.33** & 20.12*** & 1.907 & 0.600 & 0.035 & 17    & \textbf{0.48} & \textbf{2.73} & \textbf{10.98} & 1.425 & 0.850 & 0.036 \\
				Gumbel Rvine & 26    & \textbf{1.66} & \textbf{5.67*} & 34.45*** & 0.732 & 1.300 & 0.035 & 13    & \textbf{2.82*} & 6.07** & 17.52*** & 1.796 & 0.650 & 0.035 & 16    & \textbf{0.87} & \textbf{3.33} & \textbf{12.24*} & 1.471 & 0.800 & 0.035 \\
				Mixed Cvine & 25    & \textbf{1.17} & \textbf{5.46*} & 35.65*** & 0.789 & 1.250 & 0.035 & 12    & \textbf{3.77*} & 7.33** & 20.10*** & 1.878 & 0.600 & 0.035 & 19    & \textbf{0.05} & \textbf{1.91*} & \textbf{8.84} & 1.223 & 0.950 & 0.035 \\
				Mixed Dvine & 28    & \textbf{2.87*} & 6.37** & 31.55*** & 0.752 & 1.400 & 0.036 & 13    & \textbf{2.82*} & 6.07** & 17.57** & 1.782 & 0.650 & 0.035 & 19    & \textbf{0.05} & \textbf{1.91*} & 23.65*** & 1.265 & 0.950 & 0.036 \\
				Mixed Rvine & 24    & \textbf{0.76} & \textbf{5.34*} & 37.38*** & 0.893 & 1.200 & 0.036 & 12    & \textbf{3.77*} & 7.33** & 20.07*** & 2.017 & 0.600 & 0.035 & 19    & \textbf{0.05} & \textbf{1.91*} & 23.65*** & 1.256 & 0.950 & 0.035 \\
				\bottomrule
			\end{tabular}%
			\begin{tablenotes}
				\footnotesize
				\item Note: This table reports the results of VaR backtesting. NE is the number of actual exceedances. UC, CC and DQ are estimated statistics for the unconditional coverage test of \cite{kupiec1995techniques}, conditional coverage test of \cite{christoffersen1998evaluating} and Dynamic Quantile test of \cite{engle2004caviar}, respectively. AD is the mean absolute deviation between the observations and the quantile \citep{mcaleer2008forecasting}. AE is the ratio of actual over expected exceedances. AQL is the average quantile loss suggested by \cite{gonzalez2004forecasting}. In Panel A, the out-of-sample period is from January 1, 2007 to December 31, 2009. Panel B reports the result from May 3, 2005, to December 31, 2012. Bold numbers are the test statistics with p-values higher than 5\%. ***, **, * denote significance at the 1\%, 5\%, and 10\% levels, respectively.
			\end{tablenotes}
	\end{threeparttable}}
\end{table}
\end{landscape}
		\begin{ThreePartTable}
	
	\begin{TableNotes}[para,flushleft]
		\small
		\item Note: This table reports the results of ES backtesting from January 1, 2007 to December 31, 2009. CC, ER and ESR are p-values for the Conditional Calibrartion \citep{nolde2017elicitability}, Exceedance Residuals \citep{mcneil2000estimation} and Expected Shortfall Regression \citep{bayer2018regression} tests. Bold numbers are the p-values higher than 5\%. ***, **, * denote significance at the 1\%, 5\%, and 10\% levels, respectively.
	\end{TableNotes}
	
	\begin{longtable}{llllllllllllll}
		\caption{ES Back-testing (2007-2009)}\label{ESB}\\
		\toprule
    \multicolumn{1}{c}{\multirow{2}[4]{*}{Forecasting Model}} & \multicolumn{2}{l}{CC} & \multicolumn{2}{l}{ER} & \multicolumn{2}{l}{ESR} & \multicolumn{2}{l}{ESR Intercept} \\
\cmidrule{2-9}          & Simple & General & Bootstrap & Asymptotic & Bootstrap & Asymptotic & Bootstrap & Asymptotic \\
		\midrule
		\endfirsthead
		
		\multicolumn{10}{@{}l}{\ldots continued}\\
		\toprule
    \multicolumn{1}{c}{\multirow{2}[4]{*}{Forecasting Model}} & \multicolumn{2}{l}{CC} & \multicolumn{2}{l}{ER} & \multicolumn{2}{l}{ESR} & \multicolumn{2}{l}{ESR Intercept} \\
\cmidrule{2-9}          & Simple & General & Bootstrap & Asymptotic & Bootstrap & Asymptotic & Bootstrap & Asymptotic \\
		\midrule
		\endhead
		
    \multicolumn{9}{p{40.935em}}{\textit{Panel A: SR Portfolio Strategies}} \\
Student-$t$  & 0.02** & \textbf{0.08*} & \textbf{0.09*} & 0.04** & 0.01** & 0.00*** & \textbf{0.09*} & 0.00*** \\
Student-$t$ Cvine & 0.00*** & 0.01** & 0.02** & 0.00*** & 0.01** & 0.00*** & 0.02** & 0.00*** \\
Student-$t$ Dvine & 0.00*** & 0.01** & 0.01** & 0.00*** & 0.02** & 0.00*** & 0.02** & 0.00*** \\
Student-$t$ Rvine & 0.00*** & 0.03** & 0.05* & 0.01** & 0.00*** & 0.00*** & 0.01** & 0.00*** \\
Clayton & \textbf{0.12} & \textbf{0.31} & \textbf{0.23} & \textbf{0.15} & 0.05* & 0.00*** & \textbf{0.13} & 0.03** \\
Clayton Cvine & 0.02** & 0.05* & \textbf{0.07*} & 0.03** & \textbf{0.08*} & 0.00*** & \textbf{0.06*} & 0.00*** \\
Clayton Dvine & 0.01** & \textbf{0.06*} & \textbf{0.08*} & 0.03** & 0.00*** & 0.00*** & 0.01** & 0.00*** \\
Clayton Rvine & 0.01** & \textbf{0.10} & \textbf{0.12} & \textbf{0.06*} & 0.03** & 0.00*** & \textbf{0.07**} & 0.00*** \\
Frank & 0.00*** & 0.00*** & 0.00*** & 0.00*** & 0.04** & 0.00*** & 0.04* & 0.00*** \\
Frank Cvine & 0.00*** & 0.00*** & 0.00*** & 0.00*** & 0.03** & 0.00*** & 0.00*** & 0.00*** \\
Frank Dvine & 0.00*** & 0.00*** & 0.00*** & 0.00*** & 0.03** & 0.00*** & 0.00*** & 0.00*** \\
Frank Rvine & 0.00*** & 0.00*** & 0.00*** & 0.00*** & 0.00*** & 0.00*** & 0.00*** & 0.00*** \\
Joe   & 0.00*** & 0.00*** & 0.00*** & 0.00*** & 0.01** & 0.00*** & 0.00*** & 0.00*** \\
Joe Cvine & 0.01** & \textbf{0.20} & \textbf{0.30} & \textbf{0.24} & \textbf{0.08*} & 0.00*** & \textbf{0.08*} & 0.00*** \\
Joe Dvine & 0.01** & \textbf{0.09*} & \textbf{0.12} & \textbf{0.06*} & 0.00*** & 0.00*** & 0.02** & 0.00*** \\
Joe Rvine & 0.02** & \textbf{0.06*} & \textbf{0.10} & 0.04** & 0.00*** & 0.00*** & \textbf{0.06*} & 0.00*** \\
Gumbel & 0.00*** & 0.00*** & 0.00*** & 0.00*** & \textbf{0.06*} & 0.00*** & \textbf{0.08*} & 0.00*** \\
Gumbel Cvine & 0.02** & 0.03** & 0.04** & 0.01** & 0.01** & 0.00*** & 0.03** & 0.00*** \\
Gumbel Dvine & 0.00*** & \textbf{0.07*} & \textbf{0.09*} & 0.03** & 0.00*** & 0.00*** & 0.02** & 0.00*** \\
Gumbel Rvine & 0.00*** & \textbf{0.06*} & \textbf{0.10} & 0.05* & 0.00*** & 0.00*** & 0.01** & 0.00*** \\
Mixed Cvine & 0.00*** & 0.02** & 0.04** & 0.01** & 0.02** & 0.00*** & 0.02** & 0.00*** \\
Mixed Dvine & 0.01** & 0.01** & 0.01** & 0.00*** & 0.00*** & 0.00*** & 0.02** & 0.00*** \\
Mixed Rvine & 0.00*** & 0.03** & 0.04** & 0.01** & 0.00*** & 0.00*** & 0.00*** & 0.00*** \\
\multicolumn{9}{p{40.935em}}{\textit{Panel B: CVaR Portfolio Strategies}} \\
Student-$t$  & 0.04** & \textbf{0.08*} & \textbf{0.11} & 0.04** & \textbf{0.10} & 0.00*** & 0.05* & 0.02** \\
Student-$t$ Cvine & \textbf{0.11} & 0.03** & 0.01** & 0.00*** & 0.01** & 0.00*** & \textbf{0.06*} & 0.00*** \\
Student-$t$ Dvine & \textbf{0.21} & 0.03** & 0.01** & 0.00*** & 0.02** & 0.00*** & \textbf{0.06*} & 0.00*** \\
Student-$t$ Rvine & 0.04** & 0.05* & 0.04** & 0.02** & \textbf{0.06*} & 0.00*** & \textbf{0.06*} & 0.00*** \\
Clayton & 0.04** & \textbf{0.47} & \textbf{0.33} & 0.26  & \textbf{0.36} & \textbf{0.27} & \textbf{0.07*} & 0.00*** \\
Clayton Cvine & 0.00*** & 0.05* & \textbf{0.08*} & 0.03** & 0.01** & 0.00*** & 0.00*** & 0.00*** \\
Clayton Dvine & 0.01** & 0.05* & 0.05* & 0.02** & 0.02** & 0.00*** & \textbf{0.07*} & 0.00*** \\
Clayton Rvine & 0.02** & 0.04** & 0.04** & 0.01** & 0.01** & 0.00*** & \textbf{0.07*} & 0.00*** \\
Frank & \textbf{0.15} & 0.01** & 0.00*** & 0.00*** & \textbf{0.08*} & 0.00*** & 0.01** & 0.00*** \\
Frank Cvine & \textbf{0.13} & 0.02** & 0.01** & 0.00*** & 0.00*** & 0.00*** & 0.03** & 0.00*** \\
Frank Dvine & \textbf{0.13} & 0.02** & 0.00*** & 0.00*** & 0.02** & 0.00*** & 0.03** & 0.00*** \\
Frank Rvine & \textbf{0.06*} & 0.01** & 0.00*** & 0.00*** & 0.02** & 0.00*** & 0.04** & 0.00*** \\
Joe   & 0.01** & 0.00*** & 0.00*** & 0.00*** & \textbf{0.08*} & 0.00*** & 0.00*** & 0.00*** \\
Joe Cvine & 0.00*** & \textbf{0.06*} & \textbf{0.08*} & 0.03** & 0.00*** & 0.00*** & 0.05* & 0.00*** \\
Joe Dvine & 0.01** & \textbf{0.10} & \textbf{0.10} & 0.04** & 0.02** & 0.00*** & 0.05* & 0.00*** \\
Joe Rvine & 0.01** & \textbf{0.06*} & \textbf{0.14} & \textbf{0.06*} & 0.04** & 0.00*** & \textbf{0.07*} & 0.00*** \\
Gumbel & \textbf{0.17} & 0.05* & \textbf{0.09*} & 0.03** & \textbf{0.15} & 0.00*** & \textbf{0.10} & 0.01** \\
Gumbel Cvine & 0.05* & 0.02** & 0.02** & 0.00*** & 0.01** & 0.00*** & 0.02** & 0.00*** \\
Gumbel Dvine & 0.02** & \textbf{0.09*} & \textbf{0.11} & 0.05* & 0.02** & 0.00*** & \textbf{0.07*} & 0.01** \\
Gumbel Rvine & 0.01** & 0.03** & 0.03** & 0.01** & \textbf{0.15} & 0.01** & 0.02** & 0.00*** \\
Mixed Cvine & \textbf{0.08*} & 0.03** & 0.03** & 0.01** & 0.01** & 0.00*** & \textbf{0.07*} & 0.00*** \\
Mixed Dvine & \textbf{0.17} & 0.04** & 0.02** & 0.01** & 0.01** & 0.00*** & \textbf{0.06*} & 0.00*** \\
Mixed Rvine & \textbf{0.18} & 0.04** & 0.04** & 0.01** & \textbf{0.09*} & 0.00*** & \textbf{0.08*} & 0.00*** \\
\multicolumn{9}{p{40.935em}}{\textit{Panel C: GMV Portfolio Strategies}} \\
Student-$t$  & \textbf{0.23} & \textbf{0.11} & \textbf{0.11} & 0.05* & 0.01** & 0.00*** & 0.01** & \textbf{0.06*} \\
Student-$t$ Cvine & 0.00*** & 0.04** & 0.04** & 0.01** & 0.01** & 0.00*** & 0.00*** & \textbf{0.06*} \\
Student-$t$ Dvine & \textbf{0.16} & 0.03** & 0.02** & 0.00*** & 0.01** & 0.00*** & 0.00*** & 0.05* \\
Student-$t$ Rvine & \textbf{0.08*} & \textbf{0.06*} & \textbf{0.06*} & 0.02** & 0.01** & 0.00*** & 0.00*** & \textbf{0.06*} \\
Clayton & 0.00*** & 0.05* & \textbf{0.97} & \textbf{0.99} & 0.04** & 0.00*** & \textbf{0.59} & \textbf{0.63} \\
Clayton Cvine & \textbf{0.15} & \textbf{0.19} & \textbf{0.28} & \textbf{0.20} & 0.00*** & 0.00*** & 0.04** & \textbf{0.07*} \\
Clayton Dvine & \textbf{0.34} & \textbf{0.11} & \textbf{0.12} & \textbf{0.06*} & 0.01** & 0.00*** & 0.02** & \textbf{0.07*} \\
Clayton Rvine & \textbf{0.27} & \textbf{0.09*} & \textbf{0.13} & \textbf{0.06*} & 0.00*** & 0.00*** & 0.03** & 0.05* \\
Frank & 0.00*** & 0.00*** & 0.00*** & 0.00*** & \textbf{0.09*} & 0.00*** & 0.00*** & 0.02** \\
Frank Cvine & 0.01** & 0.01** & 0.00*** & 0.00*** & 0.01** & 0.00*** & 0.00*** & 0.03** \\
Frank Dvine & 0.00*** & 0.01** & 0.01** & 0.00*** & 0.01** & 0.00*** & 0.00*** & 0.03** \\
Frank Rvine & 0.00*** & 0.01** & 0.00*** & 0.00*** & 0.01** & 0.00*** & 0.00*** & \textbf{0.06*} \\
Joe   & 0.00*** & 0.00*** & 0.00*** & 0.00*** & \textbf{0.06*} & 0.00*** & 0.00*** & 0.00*** \\
Joe Cvine & \textbf{0.11} & \textbf{0.28} & \textbf{0.40} & \textbf{0.36} & 0.00*** & 0.00*** & 0.04** & \textbf{0.08*} \\
Joe Dvine & \textbf{0.30} & \textbf{0.22} & \textbf{0.22} & \textbf{0.14} & 0.00*** & 0.00*** & \textbf{0.06*} & \textbf{0.08*} \\
Joe Rvine & \textbf{0.41} & \textbf{0.08*} & \textbf{0.07**} & 0.03** & 0.01** & 0.00*** & 0.03** & \textbf{0.06*} \\
Gumbel & 0.00*** & 0.00*** & 0.00*** & 0.00*** & \textbf{0.14} & 0.00*** & 0.00*** & 0.00*** \\
Gumbel Cvine & \textbf{0.35} & \textbf{0.08*} & 0.09* & 0.04** & 0.01** & 0.00*** & 0.05* & 0.02** \\
Gumbel Dvine & \textbf{0.35} & \textbf{0.10} & 0.10  & 0.04** & 0.01** & 0.00*** & \textbf{0.07*} & 0.03** \\
Gumbel Rvine & \textbf{0.24} & \textbf{0.09*} & 0.12  & 0.05* & 0.01** & 0.00*** & \textbf{0.07*} & 0.03** \\
Mixed Cvine & \textbf{0.23} & 0.04** & 0.02** & 0.01** & 0.01** & 0.00*** & 0.00*** & \textbf{0.06*} \\
Mixed Dvine & \textbf{0.18} & \textbf{0.06*} & \textbf{0.08*} & 0.03** & 0.01** & 0.00*** & 0.00*** & 0.05* \\
Mixed Rvine & \textbf{0.18} & 0.05* & 0.04** & 0.02** & 0.01** & 0.00*** & 0.01** & \textbf{0.06*} \\
\bottomrule
		\insertTableNotes
	\end{longtable}
\end{ThreePartTable}

\newpage

		\begin{ThreePartTable}
	
	\begin{TableNotes}[para,flushleft]
		\small
		\item Note: This table reports the results of ES backtesting from May 3, 2005, to December 31, 2012. CC, ER and ESR are p-values for the Conditional Calibrartion \citep{nolde2017elicitability}, Exceedance Residuals \citep{mcneil2000estimation} and Expected Shortfall Regression \citep{bayer2018regression} tests. Bold numbers are the p-values higher than 5\%. ***, **, * denote significance at the 1\%, 5\%, and 10\% levels, respectively.
	\end{TableNotes}
	
	\begin{longtable}{llllllllllllll}
		\caption{ES Back-testing (2005-2012)}\label{ESB1}\\
		\toprule
		\multicolumn{1}{c}{\multirow{2}[4]{*}{Forecasting Model}} & \multicolumn{2}{l}{CC} & \multicolumn{2}{l}{ER} & \multicolumn{2}{l}{ESR} & \multicolumn{2}{l}{ESR Intercept} \\
		\cmidrule{2-9}          & Simple & General & Bootstrap & Asymptotic & Bootstrap & Asymptotic & Bootstrap & Asymptotic \\
		\midrule
		\endfirsthead
		
		\multicolumn{10}{@{}l}{\ldots continued}\\
		\toprule
		\multicolumn{1}{c}{\multirow{2}[4]{*}{Forecasting Model}} & \multicolumn{2}{l}{CC} & \multicolumn{2}{l}{ER} & \multicolumn{2}{l}{ESR} & \multicolumn{2}{l}{ESR Intercept} \\
		\cmidrule{2-9}          & Simple & General & Bootstrap & Asymptotic & Bootstrap & Asymptotic & Bootstrap & Asymptotic \\
		\midrule
		\endhead
		
    \multicolumn{9}{p{40.935em}}{Panel A: SR Portfolio Strategies} \\
Student-$t$  & \textbf{0.44} & \textbf{0.38} & \textbf{0.38} & \textbf{0.29} & \textbf{0.56} & \textbf{0.56} & \textbf{0.26} & \textbf{0.20} \\
Student-$t$ Cvine & \textbf{0.25} & \textbf{0.06*} & \textbf{0.09*} & 0.05* & \textbf{0.15} & \textbf{0.13} & \textbf{0.12} & 0.05* \\
Student-$t$ Dvine & \textbf{0.18} & \textbf{0.07*} & \textbf{0.10} & 0.04** & \textbf{0.15} & \textbf{0.12} & \textbf{0.10} & 0.04** \\
Student-$t$ Rvine & \textbf{0.14} & \textbf{0.09*} & \textbf{0.13} & \textbf{0.07*} & \textbf{0.14} & \textbf{0.11} & \textbf{0.11} & 0.05* \\
Clayton & \textbf{0.48} & \textbf{0.96} & \textbf{0.66} & \textbf{0.65} & \textbf{0.60} & \textbf{0.64} & \textbf{0.55} & \textbf{0.52} \\
Clayton Cvine & \textbf{0.24} & \textbf{0.37} & \textbf{0.37} & \textbf{0.29} & \textbf{0.18} & \textbf{0.25} & \textbf{0.16} & \textbf{0.13} \\
Clayton Dvine & 0.03** & \textbf{0.39} & \textbf{0.38} & \textbf{0.30} & \textbf{0.10} & \textbf{0.10} & \textbf{0.07*} & 0.05* \\
Clayton Rvine & \textbf{0.17} & \textbf{0.16} & \textbf{0.14} & \textbf{0.07*} & \textbf{0.22} & \textbf{0.17} & \textbf{0.09*} & 0.05* \\
Frank & 0.00*** & 0.00*** & 0.00*** & 0.00*** & 0.00*** & 0.00*** & 0.00*** & 0.00*** \\
Frank Cvine & 0.00*** & 0.00*** & 0.00*** & 0.00*** & 0.01** & 0.00*** & 0.04** & 0.00*** \\
Frank Dvine & 0.00*** & 0.00*** & 0.00*** & 0.00*** & 0.01** & 0.00*** & 0.04** & 0.00*** \\
Frank Rvine & 0.00*** & 0.00*** & 0.00*** & 0.00*** & 0.02** & 0.00*** & \textbf{0.09*} & 0.00*** \\
Joe   & 0.00*** & 0.00*** & 0.00*** & 0.00*** & 0.00*** & 0.00*** & 0.01** & 0.00*** \\
Joe Cvine & \textbf{0.14} & \textbf{0.47} & \textbf{0.43} & \textbf{0.35} & \textbf{0.14} & \textbf{0.22} & \textbf{0.14} & \textbf{0.11} \\
Joe Dvine & 0.05* & \textbf{0.47} & \textbf{0.38} & \textbf{0.29} & \textbf{0.16} & \textbf{0.20} & \textbf{0.10} & \textbf{0.07*} \\
Joe Rvine & \textbf{0.14} & \textbf{0.39} & \textbf{0.28} & \textbf{0.19} & \textbf{0.24} & \textbf{0.26} & \textbf{0.12} & \textbf{0.07*} \\
Gumbel & 0.00*** & 0.00*** & 0.00*** & 0.00*** & 0.01** & 0.00*** & 0.03** & 0.00*** \\
Gumbel Cvine & \textbf{0.59} & \textbf{0.45} & \textbf{0.42} & \textbf{0.32} & \textbf{0.42} & \textbf{0.51} & \textbf{0.33} & \textbf{0.31} \\
Gumbel Dvine & \textbf{0.18} & \textbf{0.45} & \textbf{0.46} & \textbf{0.40} & \textbf{0.25} & \textbf{0.29} & \textbf{0.21} & \textbf{0.14} \\
Gumbel Rvine & \textbf{0.48} & \textbf{0.30} & \textbf{0.34} & \textbf{0.24} & \textbf{0.25} & \textbf{0.34} & \textbf{0.23} & \textbf{0.17} \\
Mixed Cvine & \textbf{0.47} & \textbf{0.17} & \textbf{0.19} & \textbf{0.11} & \textbf{0.28} & \textbf{0.33} & \textbf{0.19} & \textbf{0.15} \\
Mixed Dvine & \textbf{0.27} & \textbf{0.18} & \textbf{0.31} & \textbf{0.22} & \textbf{0.20} & \textbf{0.22} & \textbf{0.18} & \textbf{0.10} \\
Mixed Rvine & \textbf{0.46} & \textbf{0.08*} & \textbf{0.12} & \textbf{0.06*} & \textbf{0.31} & \textbf{0.30} & \textbf{0.15} & \textbf{0.10} \\
\multicolumn{9}{p{40.935em}}{Panel B: CVaR Portfolio Strategies} \\
Student-$t$  & 0.00*** & \textbf{0.12} & \textbf{0.10} & \textbf{0.06*} & \textbf{0.71} & \textbf{0.75} & \textbf{0.37} & \textbf{0.31} \\
Student-$t$ Cvine & 0.00*** & \textbf{0.10} & \textbf{0.09*} & 0.05* & \textbf{0.66} & \textbf{0.69} & \textbf{0.29} & \textbf{0.23} \\
Student-$t$ Dvine & 0.00*** & \textbf{0.09*} & \textbf{0.08*} & 0.05* & \textbf{0.60} & \textbf{0.66} & \textbf{0.26} & \textbf{0.19} \\
Student-$t$ Rvine & 0.01** & \textbf{0.10} & \textbf{0.07*} & 0.05* & \textbf{0.60} & \textbf{0.63} & \textbf{0.24} & \textbf{0.20} \\
Clayton & \textbf{0.62} & \textbf{0.39} & \textbf{0.24} & \textbf{0.14} & \textbf{0.57} & \textbf{0.59} & \textbf{0.26} & \textbf{0.20} \\
Clayton Cvine & \textbf{0.57} & \textbf{0.27} & \textbf{0.20} & \textbf{0.11} & \textbf{0.58} & \textbf{0.61} & \textbf{0.24} & \textbf{0.18} \\
Clayton Dvine & \textbf{0.52} & \textbf{0.21} & \textbf{0.18} & \textbf{0.10} & \textbf{0.55} & \textbf{0.57} & \textbf{0.22} & \textbf{0.17} \\
Clayton Rvine & \textbf{0.58} & \textbf{0.31} & \textbf{0.23} & \textbf{0.13} & \textbf{0.63} & \textbf{0.62} & \textbf{0.25} & \textbf{0.19} \\
Frank & 0.00*** & 0.00*** & 0.00*** & 0.00*** & 0.05* & 0.00*** & 0.00*** & 0.05* \\
Frank Cvine & 0.01** & \textbf{0.06*} & \textbf{0.07*} & 0.04** & \textbf{0.10} & \textbf{0.09*} & 0.05* & \textbf{0.06*} \\
Frank Dvine & 0.00*** & \textbf{0.07*} & \textbf{0.09*} & 0.05* & \textbf{0.09*} & \textbf{0.09*} & 0.05* & \textbf{0.06*} \\
Frank Rvine & 0.00*** & 0.05* & \textbf{0.07*} & 0.04** & \textbf{0.12} & \textbf{0.07*} & 0.05* & \textbf{0.06*} \\
Joe   & 0.00*** & 0.00*** & 0.00*** & 0.00*** & 0.04** & 0.00*** & 0.00*** & 0.04** \\
Joe Cvine & \textbf{0.64} & \textbf{0.36} & \textbf{0.24} & \textbf{0.14} & \textbf{0.60} & \textbf{0.62} & \textbf{0.24} & \textbf{0.18} \\
Joe Dvine & \textbf{0.61} & \textbf{0.26} & \textbf{0.22} & \textbf{0.12} & \textbf{0.58} & \textbf{0.59} & \textbf{0.23} & \textbf{0.18} \\
Joe Rvine & \textbf{0.50} & \textbf{0.30} & \textbf{0.21} & \textbf{0.12} & \textbf{0.63} & \textbf{0.67} & \textbf{0.27} & \textbf{0.20} \\
Gumbel & 0.00*** & 0.00*** & 0.00*** & 0.00*** & \textbf{0.08*} & 0.00*** & \textbf{0.12} & 0.00*** \\
Gumbel Cvine & 0.02** & \textbf{0.19} & \textbf{0.15} & \textbf{0.08*} & \textbf{0.73} & \textbf{0.77} & \textbf{0.32} & \textbf{0.39} \\
Gumbel Dvine & 0.00*** & \textbf{0.15} & \textbf{0.13} & \textbf{0.07*} & \textbf{0.64} & \textbf{0.69} & \textbf{0.27} & \textbf{0.35} \\
Gumbel Rvine & 0.02** & \textbf{0.18} & \textbf{0.16} & \textbf{0.08*} & \textbf{0.69} & \textbf{0.75} & \textbf{0.31} & \textbf{0.36} \\
Mixed Cvine & 0.00*** & \textbf{0.13} & \textbf{0.12} & \textbf{0.07*} & \textbf{0.66} & \textbf{0.71} & \textbf{0.35} & \textbf{0.29} \\
Mixed Dvine & 0.01** & \textbf{0.13} & \textbf{0.13} & \textbf{0.07*} & \textbf{0.66} & \textbf{0.69} & \textbf{0.30} & \textbf{0.26} \\
Mixed Rvine & 0.00*** & \textbf{0.09*} & \textbf{0.07*} & 0.05* & \textbf{0.59} & \textbf{0.65} & \textbf{0.27} & \textbf{0.24} \\
\multicolumn{9}{p{40.935em}}{Panel C: GMV Portfolio Strategies} \\
Student-$t$  & \textbf{0.14} & \textbf{0.23} & \textbf{0.16} & \textbf{0.09*} & \textbf{0.75} & \textbf{0.78} & \textbf{0.34} & \textbf{0.38} \\
Student-$t$ Cvine & \textbf{0.41} & \textbf{0.16} & \textbf{0.15} & \textbf{0.08*} & \textbf{0.47} & \textbf{0.50} & \textbf{0.16} & \textbf{0.20} \\
Student-$t$ Dvine & \textbf{0.40} & \textbf{0.14} & \textbf{0.15} & \textbf{0.08*} & \textbf{0.44} & \textbf{0.49} & \textbf{0.15} & \textbf{0.19} \\
Student-$t$ Rvine & \textbf{0.52} & \textbf{0.19} & \textbf{0.18} & \textbf{0.10} & \textbf{0.45} & \textbf{0.51} & \textbf{0.16} & \textbf{0.20} \\
Clayton & \textbf{0.61} & \textbf{0.42} & \textbf{0.26} & \textbf{0.16} & \textbf{0.57} & \textbf{0.61} & \textbf{0.21} & \textbf{0.28} \\
Clayton Cvine & \textbf{0.56} & \textbf{0.19} & \textbf{0.16} & \textbf{0.09*} & \textbf{0.52} & \textbf{0.58} & \textbf{0.13} & \textbf{0.18} \\
Clayton Dvine & \textbf{0.53} & \textbf{0.23} & \textbf{0.19} & \textbf{0.10} & \textbf{0.50} & \textbf{0.47} & \textbf{0.15} & \textbf{0.19} \\
Clayton Rvine & \textbf{0.61} & \textbf{0.25} & \textbf{0.22} & \textbf{0.12} & \textbf{0.57} & \textbf{0.57} & \textbf{0.17} & \textbf{0.23} \\
Frank & 0.00*** & 0.00*** & 0.00*** & 0.00*** & 0.05* & 0.00*** & 0.05* & 0.00*** \\
Frank Cvine & 0.00*** & 0.04** & 0.04** & 0.04** & \textbf{0.12} & \textbf{0.06*} & \textbf{0.07*} & 0.05* \\
Frank Dvine & 0.00*** & 0.04** & 0.05* & 0.04** & \textbf{0.09*} & 0.05* & \textbf{0.06*} & 0.04** \\
Frank Rvine & 0.00*** & 0.03** & 0.04** & 0.03** & \textbf{0.11} & 0.04** & \textbf{0.09*} & 0.04** \\
Joe   & 0.00*** & 0.00*** & 0.00*** & 0.00*** & 0.05* & 0.00*** & 0.05* & 0.00*** \\
Joe Cvine & \textbf{0.51} & \textbf{0.18} & \textbf{0.14} & \textbf{0.08*} & \textbf{0.55} & \textbf{0.54} & \textbf{0.14} & \textbf{0.19} \\
Joe Dvine & \textbf{0.53} & \textbf{0.28} & \textbf{0.23} & \textbf{0.12} & \textbf{0.51} & \textbf{0.50} & \textbf{0.14} & \textbf{0.19} \\
Joe Rvine & \textbf{0.63} & \textbf{0.32} & \textbf{0.22} & \textbf{0.13} & \textbf{0.67} & \textbf{0.69} & \textbf{0.22} & \textbf{0.29} \\
Gumbel & 0.00*** & 0.00*** & 0.01** & 0.00*** & \textbf{0.12} & 0.00*** & 0.22  & 0.01** \\
Gumbel Cvine & \textbf{0.19} & \textbf{0.18} & \textbf{0.08*} & \textbf{0.14} & \textbf{0.71} & \textbf{0.72} & \textbf{0.26} & \textbf{0.31} \\
Gumbel Dvine & \textbf{0.26} & \textbf{0.15} & \textbf{0.08*} & \textbf{0.15} & \textbf{0.55} & \textbf{0.57} & \textbf{0.19} & \textbf{0.24} \\
Gumbel Rvine & \textbf{0.19} & \textbf{0.18} & \textbf{0.08*} & \textbf{0.14} & \textbf{0.67} & \textbf{0.70} & \textbf{0.23} & \textbf{0.29} \\
Mixed Cvine & \textbf{0.52} & \textbf{0.26} & \textbf{0.23} & \textbf{0.12} & \textbf{0.56} & \textbf{0.61} & \textbf{0.23} & \textbf{0.27} \\
Mixed Dvine & \textbf{0.46} & \textbf{0.18} & \textbf{0.17} & \textbf{0.09*} & \textbf{0.53} & \textbf{0.57} & \textbf{0.17} & \textbf{0.22} \\
Mixed Rvine & \textbf{0.48} & \textbf{0.19} & \textbf{0.17} & \textbf{0.09*} & \textbf{0.52} & \textbf{0.58} & \textbf{0.19} & \textbf{0.24} \\
\bottomrule
		\insertTableNotes
	\end{longtable}
\end{ThreePartTable}

\begin{singlespace}
\begin{table}
  \centering
    \caption{Regression results: Effects of modelling strategy on out-of-sample target measures, quarterly time series}\label{Reg}

{\small
\def\sym#1{\ifmmode^{#1}\else\(^{#1}\)\fi}
\begin{tabular}{lccc|ccc}
\toprule   &\multicolumn{3}{c}{\textit{Panel A: Entire period 2005-2012}}&\multicolumn{3}{c}{\textit{Panel B: Financial crisis 2007-2009}}\\
                    &\multicolumn{1}{c}{SR}&\multicolumn{1}{c}{min CVaR}&\multicolumn{1}{c}{StdDev}&\multicolumn{1}{c}{SR}&\multicolumn{1}{c}{min CVaR}&\multicolumn{1}{c}{StdDev}\\

                    &\multicolumn{1}{c}{(1)}&\multicolumn{1}{c}{(2)}&\multicolumn{1}{c|}{(3)}&\multicolumn{1}{c}{(4)}&\multicolumn{1}{c}{(5)}&\multicolumn{1}{c}{(6)}\\

\midrule

Clayton simple            &      0.0903\sym{***}&      0.0466         &     -0.0329\sym{**} &       0.179\sym{***}&       0.181\sym{***}&     -0.0272         \\
                    &      (9.14)         &      (1.27)         &     (-2.23)         &     (11.89)         &      (2.60)         &     (-1.19)         \\
Clayton Cvine        &      0.0923\sym{***}&      -0.124\sym{***}&      -0.101\sym{***}&       0.174\sym{***}&      -0.108         &      -0.129\sym{***}\\
                    &      (9.35)         &     (-3.38)         &     (-6.85)         &     (11.59)         &     (-1.55)         &     (-5.64)         \\
Clayton Dvine        &      0.0888\sym{***}&      -0.141\sym{***}&      -0.103\sym{***}&       0.168\sym{***}&      -0.173\sym{**} &      -0.138\sym{***}\\
                    &      (8.99)         &     (-3.86)         &     (-7.00)         &     (11.16)         &     (-2.47)         &     (-6.03)         \\
Clayton Rvine        &      0.0821\sym{***}&      -0.114\sym{***}&      -0.102\sym{***}&       0.157\sym{***}&      -0.117\sym{*}  &      -0.131\sym{***}\\
                    &      (8.31)         &     (-3.11)         &     (-6.95)         &     (10.45)         &     (-1.67)         &     (-5.75)         \\
Frank simple              &      0.0973\sym{***}&     -0.0186         &     -0.0295\sym{**} &       0.179\sym{***}&    0.000607         &     -0.0342         \\
                    &      (9.85)         &     (-0.51)         &     (-2.00)         &     (11.89)         &      (0.01)         &     (-1.50)         \\
Frank Cvine          &      0.0929\sym{***}&      -0.148\sym{***}&     -0.0997\sym{***}&       0.172\sym{***}&      -0.167\sym{**} &      -0.134\sym{***}\\
                    &      (9.41)         &     (-4.04)         &     (-6.76)         &     (11.45)         &     (-2.40)         &     (-5.85)         \\
Frank Dvine          &      0.0909\sym{***}&      -0.149\sym{***}&     -0.0990\sym{***}&       0.171\sym{***}&      -0.176\sym{**} &      -0.133\sym{***}\\
                    &      (9.21)         &     (-4.08)         &     (-6.72)         &     (11.39)         &     (-2.53)         &     (-5.81)         \\
Frank Rvine          &      0.0898\sym{***}&      -0.152\sym{***}&      -0.100\sym{***}&       0.161\sym{***}&      -0.178\sym{**} &      -0.132\sym{***}\\
                    &      (9.10)         &     (-4.16)         &     (-6.81)         &     (10.72)         &     (-2.56)         &     (-5.80)         \\
Gumbel simple             &       0.101\sym{***}&     -0.0555         &     -0.0497\sym{***}&       0.191\sym{***}&    -0.00587         &     -0.0332         \\
                    &     (10.26)         &     (-1.52)         &     (-3.37)         &     (12.70)         &     (-0.08)         &     (-1.45)         \\
Gumbel Cvine         &      0.0939\sym{***}&      -0.123\sym{***}&      -0.100\sym{***}&       0.173\sym{***}&      -0.117\sym{*}  &      -0.132\sym{***}\\
                    &      (9.51)         &     (-3.37)         &     (-6.79)         &     (11.51)         &     (-1.67)         &     (-5.77)         \\
Gumbel Dvine         &      0.0949\sym{***}&      -0.150\sym{***}&      -0.102\sym{***}&       0.173\sym{***}&      -0.161\sym{**} &      -0.135\sym{***}\\
                    &      (9.62)         &     (-4.11)         &     (-6.89)         &     (11.49)         &     (-2.31)         &     (-5.90)         \\
Gumbel Rvine         &      0.0898\sym{***}&      -0.128\sym{***}&      -0.101\sym{***}&       0.162\sym{***}&      -0.132\sym{*}  &      -0.131\sym{***}\\
                    &      (9.10)         &     (-3.50)         &     (-6.88)         &     (10.76)         &     (-1.90)         &     (-5.74)         \\
historical          &     -0.0402\sym{***}&     -0.0816\sym{**} &      -0.130\sym{***}&     -0.0486\sym{***}&     -0.0118         &      -0.133\sym{***}\\
                    &     (-4.07)         &     (-2.23)         &     (-8.82)         &     (-3.23)         &     (-0.17)         &     (-5.83)         \\
Joe simple                &      0.0970\sym{***}&     -0.0314         &     -0.0183         &       0.182\sym{***}&     -0.0255         &     -0.0346         \\
                    &      (9.83)         &     (-0.86)         &     (-1.24)         &     (12.06)         &     (-0.37)         &     (-1.51)         \\
Joe Cvine            &      0.0891\sym{***}&      -0.126\sym{***}&      -0.102\sym{***}&       0.168\sym{***}&      -0.108         &      -0.129\sym{***}\\
                    &      (9.03)         &     (-3.44)         &     (-6.93)         &     (11.18)         &     (-1.55)         &     (-5.63)         \\
Joe Dvine            &      0.0867\sym{***}&      -0.138\sym{***}&      -0.103\sym{***}&       0.165\sym{***}&      -0.172\sym{**} &      -0.144\sym{***}\\
                    &      (8.78)         &     (-3.76)         &     (-7.00)         &     (10.97)         &     (-2.47)         &     (-6.28)         \\
Joe Rvine            &      0.0859\sym{***}&     -0.0948\sym{***}&      -0.106\sym{***}&       0.160\sym{***}&     -0.0795         &      -0.134\sym{***}\\
                    &      (8.70)         &     (-2.59)         &     (-7.17)         &     (10.64)         &     (-1.14)         &     (-5.86)         \\
mixed Cvine          &      0.0918\sym{***}&      -0.147\sym{***}&      -0.100\sym{***}&       0.171\sym{***}&      -0.159\sym{**} &      -0.139\sym{***}\\
                    &      (9.30)         &     (-4.03)         &     (-6.81)         &     (11.37)         &     (-2.28)         &     (-6.09)         \\
mixed Dvine          &      0.0955\sym{***}&      -0.149\sym{***}&     -0.0977\sym{***}&       0.176\sym{***}&      -0.173\sym{**} &      -0.133\sym{***}\\
                    &      (9.67)         &     (-4.08)         &     (-6.63)         &     (11.66)         &     (-2.49)         &     (-5.80)         \\
mixed Rvine          &      0.0900\sym{***}&      -0.143\sym{***}&      -0.101\sym{***}&       0.163\sym{***}&      -0.161\sym{**} &      -0.134\sym{***}\\
                    &      (9.11)         &     (-3.90)         &     (-6.85)         &     (10.84)         &     (-2.31)         &     (-5.88)         \\
Student-$t$ simple           &      0.0939\sym{***}&     -0.0983\sym{***}&     -0.0977\sym{***}&       0.175\sym{***}&     -0.0840         &      -0.132\sym{***}\\
                    &      (9.51)         &     (-2.68)         &     (-6.62)         &     (11.64)         &     (-1.20)         &     (-5.77)         \\
Student-$t$ Cvine       &      0.0948\sym{***}&      -0.161\sym{***}&     -0.0991\sym{***}&       0.175\sym{***}&      -0.189\sym{***}&      -0.134\sym{***}\\
                    &      (9.60)         &     (-4.39)         &     (-6.72)         &     (11.63)         &     (-2.71)         &     (-5.87)         \\
Student-$t$ Dvine       &      0.0951\sym{***}&      -0.162\sym{***}&     -0.0991\sym{***}&       0.175\sym{***}&      -0.199\sym{***}&      -0.134\sym{***}\\
                    &      (9.64)         &     (-4.42)         &     (-6.72)         &     (11.62)         &     (-2.85)         &     (-5.87)         \\
Student-$t$ Rvine       &      0.0923\sym{***}&      -0.140\sym{***}&     -0.0995\sym{***}&       0.165\sym{***}&      -0.167\sym{**} &      -0.132\sym{***}\\
                    &      (9.35)         &     (-3.82)         &     (-6.75)         &     (10.99)         &     (-2.39)         &     (-5.78)         \\
Constant            &       0.278\sym{***}&       0.657\sym{***}&       0.484\sym{***}&     -0.0267\sym{**} &       1.906\sym{***}&       0.947\sym{***}\\
                    &     (26.85)         &     (17.13)         &     (31.30)         &     (-2.09)         &     (32.20)         &     (48.82)         \\

\midrule
Observations        &         775         &         775         &         775         &         300         &         300         &         300         \\
\(R^{2}\)           &       0.947         &       0.986         &       0.990         &       0.956         &       0.990         &       0.995         \\
\bottomrule
\multicolumn{7}{p{0.7\textwidth}}{\footnotesize Note: \textit{t} statistics in parentheses, quarterly time dummies included. \sym{*} \(p<0.10\), \sym{**} \(p<0.05\), \sym{***} \(p<0.01\). Reference portfolio: EQW}\\
\end{tabular}
}
\end{table}
\end{singlespace}

\end{document}